\documentclass[a4paper,11pt]{article}
\pdfoutput=1 

\usepackage{jheppub}

\usepackage[mmddyyyy,hhmmss]{datetime}

\usepackage{amsmath,graphicx,amssymb,subfigure,bbm,comment,mathtools}
\usepackage[all]{xy}
\usepackage{slashed}

\begin{document}


\title{Giant Gravitons in the Schr\"{o}dinger holography}

\author[a]{George Georgiou}
\author[a,b]{and Dimitrios Zoakos}

\affiliation[a]{Department of Physics, National and Kapodistrian University of Athens, 15784 Athens, Greece}
\affiliation[b]{Hellenic American University, 436 Amherst st, Nashua, NH 03063 USA}

\emailAdd{ggeo@phys.uoa.gr}
\emailAdd{zoakos@gmail.com}

\abstract{\today \, at \currenttime} 

\abstract{
We construct and study new giant graviton configurations in the framework of the non-supersymmetric 
Schr\"{o}dinger holography. 
We confirm in the original Schr\"{o}dinger spacetime, the picture discovered previously
in the pp-wave limit  of the geometry, namely that it is the giant graviton that becomes the energetically favored 
stable configuration compared to the point graviton one.
Furthermore, there is a critical value of the deformation above which the point graviton disappears from the spectrum.
The former fact leads also to the possibility of tunnelling from the point graviton to the giant graviton configuration. 
We calculate, explicitly, the instanton  solution and its corresponding action which gives a measure of the tunnelling probability. 
Finally, we evaluate holographically the three-point correlation function of two giant gravitons and one dilaton mode as a 
function of the Schr\"{o}dinger invariant.}


\maketitle
\flushbottom


\section{Introduction}

An important and efficient tool in exploring a gravity background (especially one that comes as a continuous deformation of the 
${\cal N}=4$ SYM) is the inclusion of brane probes and the subsequent study of their dynamics. 
A prominent example in this class of configurations is a D3-brane that wraps a three-sphere inside the five-sphere 
of the celebrated $AdS_5 \times S^5$ geometry and it known as the giant graviton \cite{McGreevy:2000cw}. 
It has the same quantum numbers as the point particle and a nonzero angular momentum along the equator 
of the internal space. Stability against shrinking is guaranteed by the RR repulsion but also the perturbative stability analysis 
around the giant graviton solution has been performed in \cite{Das:2000st}.
When the three-sphere that the giant wraps is inside the $AdS$ part of the space the configuration to be 
realized is the dual giant graviton, with similar properties \cite{Grisaru:2000zn,Hashimoto:2000zp}.

While in the original $AdS_5 \times S^5$ geometry (and depending on the value of the conjugate angular momentum) the point graviton 
is either energetically favored or at most degenerate with respect to the giant graviton, the situation becomes intricate and interesting when the 
geometry gets deformed. 
When the deformation is purely on a sphere, as in the marginally deformed backgrounds, 
there is a variety of behaviors but in none of them the energetically favored solution is the giant graviton.
More specifically, there are cases that the degeneracy that is 
inherited by the parent undeformed background is intact \cite{Pirrone:2006iq,Imeroni:2006rb}, 
cases that the degeneracy is lifted in favor of the point graviton  \cite{deMelloKoch:2005jg,Avramis:2007wb}
and cases where the initially energetically favored point graviton becomes even more pronounced \cite{deMelloKoch:2005jg}.  
However, when the deformation involves directions  of the anti de Sitter space, as it happens in the cases of the non-commutative 
or the dipole deformation, there are situations where the energetically favored solution is the giant graviton 
(see \cite{Huang:2007th} for the non-commutative example and \cite{Georgiou:2020qnh} for the pp-wave geometry of certain
dipole deformed theories).  

There are several interesting examples of gauge/gravity dualities in which the original AdS/CFT scenario \cite{Maldacena:1997re} is deformed. One such example is the duality between a certain spacetime with Schr\"{o}dinger symmetry and its dual dipole-deformed non-relativistic conformal 
field theory \cite{Maldacena:2008wh}. Compared to the original AdS/CFT duality very few observables have been calculated in the  Schr\"{o}dinger/dipole CFT version of the correspondence. In particular, in \cite{Fuertes:2009ex} and \cite{Volovich:2009yh}  two and three, as well as   $n$-point correlation functions of scalar operators were calculated by employing the gravity side of the correspondence. It is important to mention that all the operators involved in these correlators correspond to point-like strings propagating in the $Sch_5 \times S^5$ background. In
\cite{Georgiou:2017pvi} extended dyonic giant magnon and spike solutions were constructed and their dispersion relations 
were derived.\footnote{Giant magnon solutions in other deformed geometries like the $\beta$-deformed theories have been found in \cite{Chu:2006ae}.}
A complementary study with finite size corrections of those classical solutions was presented in \cite{Zoakos:2020gyb}.\footnote{Giant magnon-like solutions with a different dispersion relation were found in
\cite{Ahn:2017bio}, while giant magnons and spiky strings on
$Sch_5 \times T^{1,1}$ and the corresponding dispersion relations were found in  
\cite{Golubtsova:2020fpm}.} 
The existence of the aforementioned solutions 
is in complete agreement with the fact that the theories involved in the  Schr\"{o}dinger/dipole CFT duality remain integrable despite the presence of the deformation. 
In the same work \cite{Georgiou:2017pvi}  an exact, in the 't Hooft coupling $\lambda$,
expression for the dimensions of certain gauge invariant operators dual to the giant magnons was
conjectured. Subsequently, this conjecture was further improved in \cite{Georgiou:2019lqh} in such a way that it is in perfect agreement with the
string spectrum in the pp-wave limit of the Schr\"{o}dinger spacetime.
Furthermore, agreement between this expression and the one-loop anomalous dimension of BMN-like operators was found  in the large $J$ limit providing, thus,  further evidence in favor of the correspondence.
On the field theory side, the one-loop spectrum of operators belonging in a $SL(2)$ closed sub-sector of the null dipole-deformed CFT has been studied in \cite{Guica:2017jmq}.  The authors found agreement between the one-loop anomalous dimensions of certain long operators and the energies of the dual string solutions (see \cite{Ouyang:2017yko}, as well).

Subsequently, by utilising the Schr\"{o}dinger background, the  three-point functions involving two {\it heavy}  and one {\it light} operator were calculated holographically in \cite{Georgiou:2018zkt}.
The {\it heavy} states were 
generalizations either of the giant magnon or spike solutions constructed in \cite{Georgiou:2017pvi} while the {\it light} operator was chosen to
be one of the dilaton modes. These
results for the three-point functions are the first in the literature in which
the {\it heavy} states participating in three-point correlation functions
are described by extended string solutions.
The aforementioned results are in complete agreement with the form of the correlator dictated by non-relativistic conformal invariance and provide us with  the leading term of the correlators in the large $\lambda$ expansion.
Finally, pulsating strings solutions in the Schr\"{o}dinger background were recently derived in \cite{Dimov:2019koi,Golubtsova:2020mjn} while the holographic Fisher information metric in Schr\"odinger spacetime was studied in \cite{Dimov:2020fzi}.

Very recently, giant graviton solutions in the pp-wave limit of the Schr\"{o}dinger geometry
were constructed in \cite{Georgiou:2020qnh}. These solutions have the remarkable property that the deformation breaks the degeneracy
between the point and giant graviton solutions in favour of the giant graviton which becomes the energetically favoured stable solution.  
The focus of the current paper is twofold. On one hand, it is  to identify the classical configurations describing giant gravitons in the Schr\"{o}dinger background before taking the pp-wave limit and on the other 
 to use these solutions in order to probe the  holography in Schr\"{o}dinger spacetimes. 
This second goal will be accomplished through the calculation of the three-point correlation functions of the dilaton modes and two 
``heavy" operators. The giant gravitons will serve as the gravity duals of the heavy states 
and the result of the holographic computation will provide us with  the leading term  in the large $\lambda$ expansion 
of the the three-point correlators.

The plan of the paper is as follows: In section \ref{section-Giant-Graviton}, we revisit the type IIB supergravity solution of the 
Schr\"{o}dinger background and present a consistent giant graviton ansatz. The Hamiltonian of the system is a function of 
two variables: the deformation parameter and the conjugate momentum with respect to one of the internal angles of the five-sphere.
We investigate the behavior of the system either fixing the conjugate momentum and varying the deformation parameter or 
vice versa and extract a plethora of interesting observations.
The importance of the ansatz we introduce is that the zero deformation limit of our Hamiltonian leads smoothly 
to the  Hamiltonian of the undeformed $AdS_5 \times S^5$  that appears in \cite{McGreevy:2000cw,Grisaru:2000zn}.  
In section \ref{section:instantons} we focus the attention to 
potential tunnelling effects between the two graviton solutions and calculate instanton transitions, in that part of the parametric space where  the giant graviton 
is the energetically favored solution. Emphasis is put on a novel instanton solution that arrises when the deformation 
parameter and the conjugate momentum are related through a constraint in such a way that, despite the presence of a deformation, the energies of the two gravitons, the point and the giant graviton, are equal.  
In section \ref{section:3-point-function}, we compute holographically the three-point correlation function of the 
dilaton modes and two ``heavy" operators whose gravity duals are the giant gravitons, in  the special case where the giant and the point graviton 
are degenerate. The result is in complete agreement with the constraints put by conformal invariance.
In section \ref{section:conclusions} we conclude the paper and identify potential future directions for 
endeavors in the framework of Schr\"{o}dinger holography.

\section{Giant Gravitons in a Schr\"{o}dinger geometry}
\label{section-Giant-Graviton}

The starting point for the analysis we will present in this paper is the expressions for the supergravity solution of the 10d 
$Sch_5 \times S^5$ geometry in global coordinates \cite{Alishahiha:2003ru,Guica:2017jmq}
\begin{eqnarray} \label{initial-Sch}
\frac{ds^2}{L^2} & = & - \left(1\, + \, \frac{\mu^2}{Z^4} \right) dT^2 +
\frac{1}{Z^2} \, \Bigg(2 \, dT \, dV + dZ^2 + d{\vec X}^2 - {\vec X}^2  dT^2 \Bigg) + ds^2_{S^5}
\nonumber \\ [7pt]
ds^2_{\rm S^5}& = & \left(d\chi+\omega\right)^2+ ds^2_{\rm \mathbb{C}P^2}
\quad {\rm \&} \quad
ds^2_{\rm \mathbb{C}P^2}= \frac{d\rho^2}{1-\rho^2}+\rho^2\, 
\Bigg[\Sigma_1^2+\Sigma_2^2+\left(1-\rho^2\right)\,\Sigma_3^2\Bigg]
\end{eqnarray}
where we have written the metric of the five-sphere as an $S^1$-fibration over the $\rm {\mathbb C}P^2$ and
the quantities $\Sigma_1$, $\Sigma_2$ and $\Sigma_3$ are defined as follows
\begin{eqnarray}
\Sigma_1&\equiv& \frac{1}{2}(\cos\psi\, d\theta - \sin\psi\sin\theta\, d\phi)
\qquad
\Sigma_2\equiv \frac{1}{2}(\sin\psi\, d\theta + \cos\psi\sin\theta\, d\phi)
\nonumber \\[5pt]
\Sigma_3&\equiv& \frac{1}{2}(d\psi - \cos\theta\, d\phi)
\qquad {\rm and} \qquad
\omega \equiv \rho^2\, \Sigma_3\,. 
\end{eqnarray}
The metric is accompanied with a magnetic field 
\begin{equation}
\frac{B_2}{L^2} = \frac{\mu}{Z^2}\,dT \wedge \left(d\chi + \omega \right)
\end{equation}
and a RR five-form
\begin{equation}
\frac{F_5}{L^4} = \frac{4}{Z^5}\,dT \wedge dV \wedge d{\vec X} \wedge dZ \, + \, 
4 \, \rho^3 \, d\rho \wedge \Sigma_1 \wedge \Sigma_2 \wedge \Sigma_3 \wedge d\chi  \, . 
\end{equation}

In order to describe Giant Graviton solutions in the Schr\"{o}dinger background \eqref{initial-Sch}, 
we have to consider the action of a probe D3-brane. This action consists of the sum of the Dirac-Born-Infeld (DBI) 
term and of the Wess-Zumino (WZ) term 
\begin{equation} \label{D3-action}
S_{{\rm D}3} = - T_3 \int d^4 \xi  \, e^{-\Phi} \sqrt{\big. - \det {\cal P}\Big[g+B+2 \pi \alpha' F \Big] } +
 T_3 \int \sum_q {\cal P}\Big[ A_q \wedge e^{B+2 \pi \alpha' F} \Big]
\end{equation}
where ${\cal P}$ denotes the pullback of the different spacetime fields 
on the brane worldvolume and $T_3$ is the tension of the D3-brane.

We consider a D3-brane probe that extends along the following directions 
\begin{equation}
\xi_0 = \tau \, ,\quad 
\xi_1 = \theta \, ,\quad 
\xi_2 = \phi \quad \& \quad 
\xi_3 = \psi
\end{equation}
and the ansatz for the rest of the coordinates is as follows
\begin{equation} \label{D3-ansatz}
T = \kappa \, \tau \, ,\quad 
V = \nu \, \tau \, ,\quad 
\vec{X} = 0  \, ,\quad
Z = Z_0 \, ,\quad
\rho = \frac{\rho_0}{L}
\quad \& \quad
\chi= \omega \, \tau \, . 
\end{equation}
Notice that $\rho_0$ is the size of the graviton in the internal space.

Substituting the ansatz \eqref{D3-ansatz} in \eqref{D3-action} and integrating the spatial coordinates of the 
world volume, we obtain the following expressions for the on-shell action and the Lagrangian
\begin{equation} \label{Lagrangian}
S_{{\rm D}3} = \int d\tau \, L_{{\rm D}3} \quad {\rm with} \quad 
L_{{\rm D}3} = -\, \frac{N}{L^4} \, 
\Bigg[ \rho_0^3 \, \sqrt{1 + \Gamma - \left(L^2 - \rho_0^2\right) \, 
\left(\omega^2 - \frac{\Delta^2}{L^2}\right)} - \rho_0^4 \, \omega \Bigg]
\end{equation}
where, to simplify notation, we have introduced the following definitions of constants
\begin{equation}
\kappa = \frac{1}{L} \ , \quad 
\Gamma = - \frac{2 \, L}{Z_0^2} \, \nu \ , \quad 
\Delta = \frac{\mu}{Z_0^2}
\quad \& \quad
2 \, \pi^2 \, T_3 = \frac{N}{L^4} \, . 
\end{equation}
The conjugate momenta to $\chi$ and $V$ are  given by
\begin{equation} \label{conj_momentum}
J_1 = {\partial L_{{\rm D}3} \over \partial \omega}  
\quad \& \quad 
J_2 = {\partial L_{{\rm D}3} \over \partial \Gamma}\, . 
\end{equation}
An important comment is in order. In contradistinction to the giant graviton ansatz in the pp-wave Schr\"{o}dinger
geometry \cite{Georgiou:2020qnh}, in the current case there are three conserved quantities (namely $E$, $J_1$ 
and $J_2$) instead of two ($E$ and $J_1$ in the pp-wave case). The need for a generalization of the previous giant graviton ansatz is coming for the following reason. 
It is known that the ansatz \eqref{D3-ansatz} satisfies the equations of motion only at the extrema of the on-shell action 
(see also the discussion in  \cite{Georgiou:2020qnh}). 
Following this reasoning, we calculated the equations of motion for the coordinates that are orthogonal to the 
brane worldvolume directions (namely $T, V, {\vec X}, Z, \rho$ and $\chi$) and indeed we realized that 
the equations of motion for $Z$ and $\rho$ are not automatically satisfied. 
The equation of motion for $\rho$ is satisfied when we impose that the derivative of the Lagrangian 
with respect to $\rho$ vanishes.
However, the equation of motion for $Z$ is not satisfied unless we fix the value of $J_2$. 
{\it Turning on} an extra coordinate dependence in the giant graviton ansatz (that for $V$) is necessary and gives us the extra freedom 
that is needed in order for all the equations of motion to be satisfied at the extrema of the Lagrangian.  
For this to happen the conjugate momentum $J_2$ should acquire the following value 
\begin{equation} \label{Vr0}
J_2 \, = \, \frac{N}{2 \, L} \, \frac{V(\rho_0)}{\sqrt{1 - \Delta^2 \left(1-\frac{\rho_0^2}{L^2}\right)}} 
\quad {\rm with} \quad
\Big(V(\rho_0)\Big)^2 \, = \, 
\frac{\rho_0^6}{L^6}+\frac{\Big({J_1 \over N} - \frac{\rho_0^4}{L^4}\Big)^2}{1-\frac{\rho_0^2}{L^2}} \, . 
\end{equation}
The corresponding Hamiltonian $H= J_1 \, \omega+J_2\,  \Gamma-L_{{\rm D}3}$ for the giant graviton solution in the Schr\"{o}dinger background becomes
\begin{equation} \label{Hamiltonian}
H \, = \, \frac{N}{L} \, \frac{V(\rho_0)}{\sqrt{1 - \Delta^2 \left(1-\frac{\rho_0^2}{L^2}\right)}} \, . 
\end{equation}
Taking the limit of zero deformation (i.e. $\Delta \rightarrow 0$) the denominator becomes the unit and we recover 
the Hamiltonian of the ${\cal N}=4$ giant graviton on the sphere \cite{McGreevy:2000cw,Grisaru:2000zn,Hashimoto:2000zp}.
The quantity inside the square root in the denominator of \eqref{Hamiltonian} should be always positive, 
and this sets an upper bound for the value of $\Delta$.
The range of values for the deformation parameter is 
\begin{equation}
0 \, \le \, \Delta \, \le \, \left(1-\frac{\rho_0^2}{L^2}\right)^{-1/2} \,  . 
\end{equation}

\begin{figure}[ht] 
   \centering
   \includegraphics[width=7.3cm]{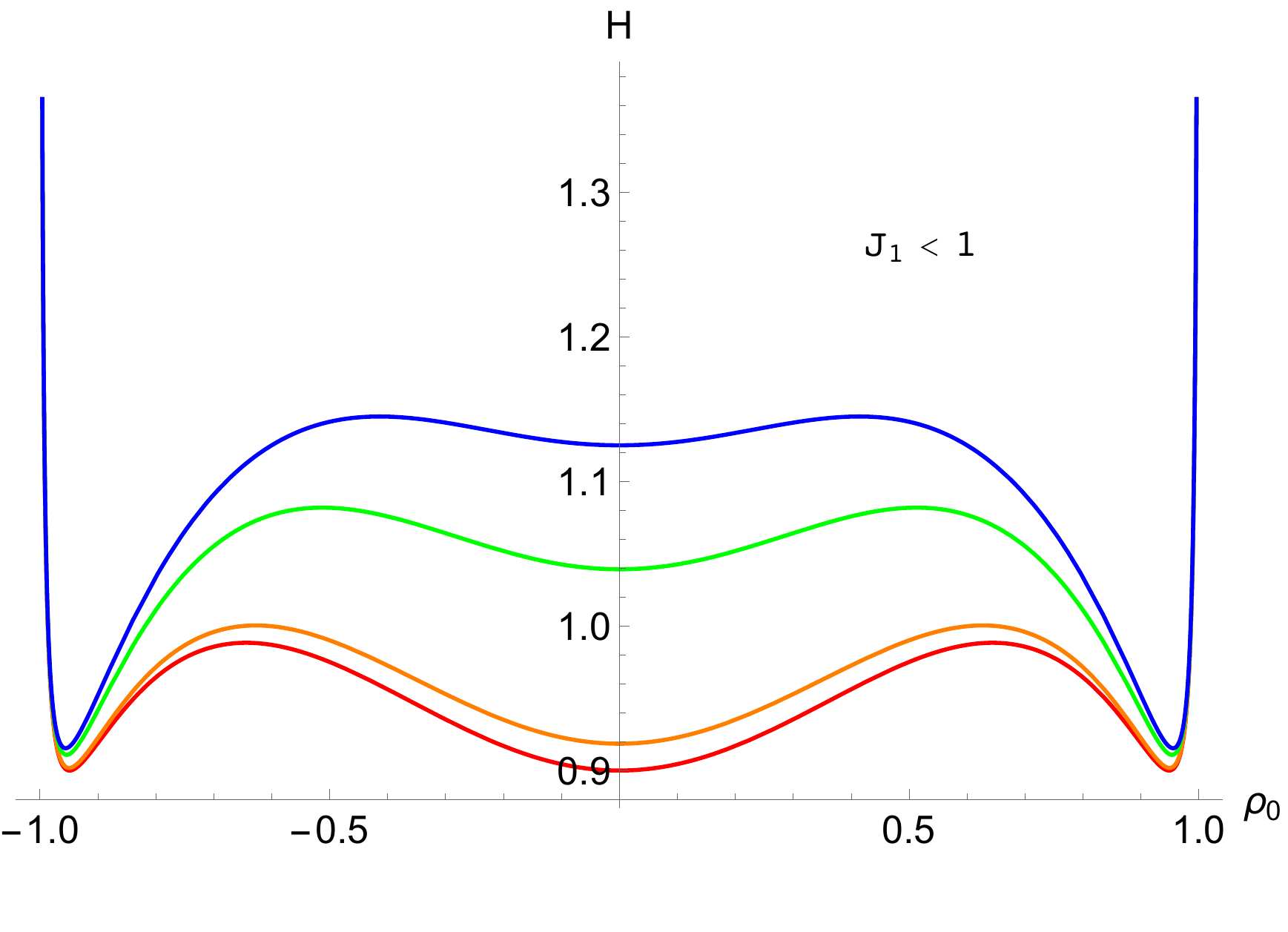}
   \hspace{0.2cm}
    \includegraphics[width=7.3cm]{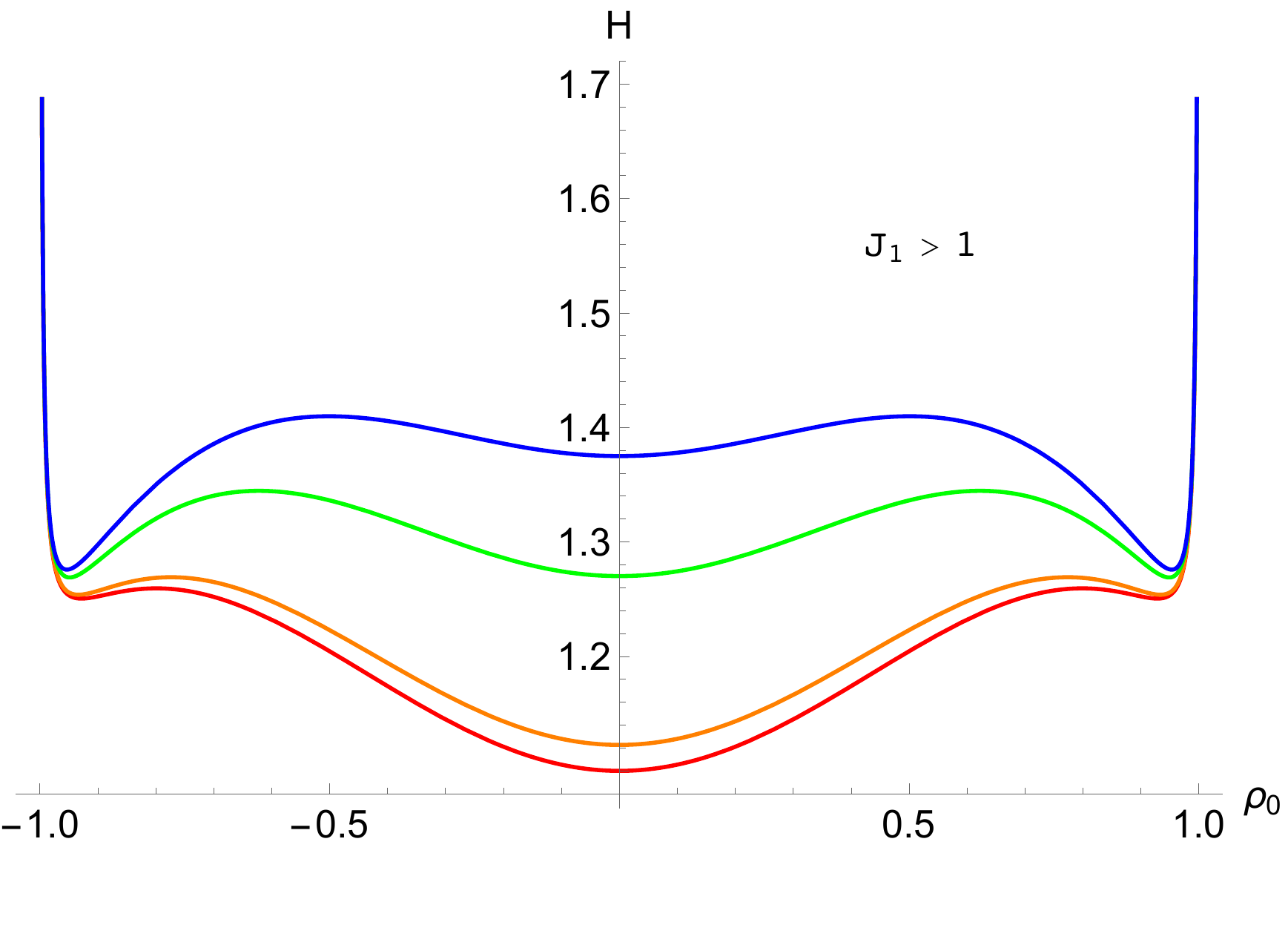}
    \caption{Hamiltonian \eqref{Hamiltonian} as a function of the graviton's size $\rho_0$,
     for different values of the deformation $\Delta$.
     The left (right) plot is for values of $J_1<1$ ($J_1>1$) and in units $N=L=1$.
     The correspondence between color and values of $\Delta$ for both plots of this figure is 
     Red $\Rightarrow$ $\Delta = 0$, Orange $\Rightarrow$ $\Delta = 0.2$, Green $\Rightarrow$ $\Delta = 0.5$
     \& Blue $\Rightarrow$ $\Delta = 0.6$. For both graphs when $\Delta >1/ \sqrt{2}$ the point graviton becomes a maximum and the potential acquairs the shape of a Mexican-hat with the giant gravitons being the only minima. }
   \label{fig:Hamiltonian}
\end{figure}

In figure \ref{fig:Hamiltonian} we plot the Hamiltonian \eqref{Hamiltonian} as a function of the graviton's size $\rho_0$,
for different values of the deformation $\Delta$ and of the conjugate momentum $J_1$.\footnote{We follow the presentation 
style that was introduced in \cite{Grisaru:2000zn} and allow $\rho_0$ to take negative values. The range of negative values
corresponds to the D3-brane expanding into the five-sphere with the angular coordinate having 
the opposite orientation. Since the potential \eqref{Hamiltonian} is even under $\rho_0 \rightarrow - \rho_0$ the negative 
$\rho_0$ configuration is identical to the positive $\rho_0$ configuration.} In all expressions and figures below we have chosen units such that the radius of the 5-sphere is $L=1$ and we have also set $N=1$, 
that is whereever one sees $J_1$ it is $J_1/N$ that should appear. 
On the left plot, we fix $J_1$ to a value that is less than one (more specifically $J_1=0.9$) and increase the value 
of the deformation parameter $\Delta$. For $\Delta=0$ we recover the ${\cal N}=4$ result, where the energy of the point 
and of the giant graviton are the same. Turning on the deformation parameter the initial degeneracy between the 
three extrema disappears and remarkably, the energetically favored solution (the one having less energy) is the giant graviton. 
This behavior is in qualitative agreement with the giant graviton calculation in the pp-wave Schr\"{o}dinger background 
\cite{Georgiou:2020qnh}. Keep increasing the value of $\Delta$, we will continue having three extrema, but the extremum 
at the position $\rho_0=0$ will change from a minimum to a maximum at the value $\Delta = 1/\sqrt{2}$. At exactly 
this value for $\Delta$ the second derivative of the Hamiltonian \eqref{Hamiltonian} with respect to $\rho_0$ 
at $\rho_0 =0$ will change sign from positive to negative and if one keeps increasing the deformation parameter there is a value of $\Delta=1$ above which the point graviton will cease to exist.
Notice that the behavior of the Hamiltonian as a function of the graviton's size $\rho_0$ will remain qualitatively unaltered
if we change the value of $J_1$, as long as we stay in the range of $J_1<1$.

On the right plot of figure \ref{fig:Hamiltonian}, we fix $J_1$ to a value that is above one (more specifically $J_1=1.1$) 
and increase the value of the deformation parameter $\Delta$. For $\Delta=0$ we have three extrema and the 
extremum at $\rho_0=0$ is the energetically favored solution, which is the ${\cal N}=4$ result. Increasing the value of 
$\Delta$, increases the value of the energy at the $\rho_0=0$ extremum and decreases the values of the other two extrema. 
As a result of this, there is a critical value of the deformation parameter $\Delta$ for which the three extrema have the 
same energy. This value depends on the value of $J_1$ and in figure  \ref{DeltavsJ} we depict the relation between the 
critical  $\Delta$ and $J_1$. In the next section we will analyze in detail this special class of solutions. 
In this region of $\Delta< \Delta_{\rm crit}$ the point graviton solution is energetically favored with respect to the giant 
graviton solution. 
The behavior of the Hamiltonian for values of the deformation parameter $\Delta$ above the critical value is
identical to the behavior we already described for $J_1<1$. More specifically, the giant graviton becomes the 
energetically favored solution and for values above $\Delta = 1/\sqrt{2}$ the extremum at the point graviton position 
turns from a minimum to a maximum and for $\Delta > 1$ the point graviton is not any more part of the spectrum (see also the discussion in \cite{Georgiou:2020qnh}). . 

.

\begin{figure}[ht] 
   \centering
   \includegraphics[width=8cm]{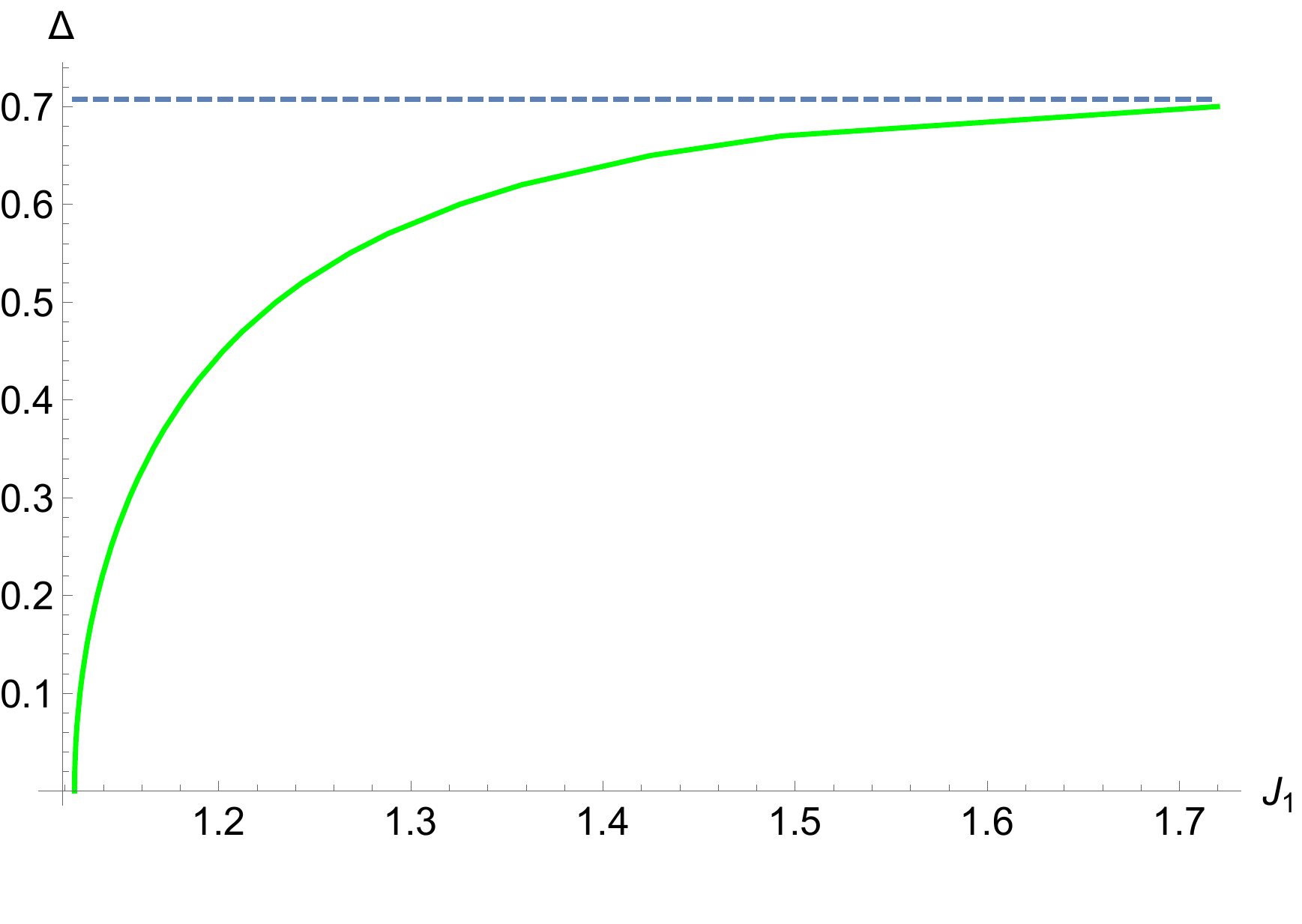}
    \caption{Plot of the conjugate momentum $J_1$ for different values of the deformation parameter $\Delta$ (all 
    of them below the value $\Delta = 1/\sqrt{2}$). Above the green line there is a giant graviton solution while below the 
    green line there is only the point graviton solution.}
   \label{DeltavsJ_gg_disappearance}
\end{figure}

Instead of fixing the conjugate momentum $J_1$ and varying the deformation parameter $\Delta$, we could have done the 
inverse. Such an analysis shows that keeping the value of the deformation parameter fixed and below 
$\Delta = 1/\sqrt{2}$, that is in the region where the point graviton is  still a minimum,
and increasing the value of the conjugate momentum beyond $J_1=1$, there is a critical value of $J_1$ 
above which the giant graviton solution completely disappears. \footnote{Notice that this happens only for values $J_1>1.125$. For $J_1\leq1.125$ the giant graviton solutions are always present, as metastable states at worst.} At this value of $J_1$ both the first and the second 
derivative of the Hamiltonian with respect to $\rho_0$ at the giant graviton position vanish. In figure 
\ref{DeltavsJ_gg_disappearance} we have plotted this
critical value of $J_1$ as a function of the deformation parameter $\Delta$.  In the undeformed ${\cal N} =4$ case, 
the critical value of the conjugate momentum is $J_1=1.125$ \cite{Grisaru:2000zn}. 
Notice that if the value of the deformation parameter 
is above $\Delta = 1/\sqrt{2}$ then the extrema corresponding to the giant graviton solution never disappear. 
This is related to the fact that  for values of the deformation parameter above $\Delta = 1/\sqrt{2}$ the extremum at the point graviton 
position turns from a minimum to a maximum.

\section{Instanton Transitions}\label{section:instantons}

The analysis of the previous section has uncovered two brane configurations that in the framework of a 
Schr\"{o}dinger background describe a graviton state with angular momentum $J_1$: 
from one side it is the point-like graviton and 
from the other side the giant graviton, a configuration that consists of a spherical D3-brane extending into the five-sphere. 
Interestingly, even though the five-sphere is not deformed in the Schr\"{o}dinger deformation of the parent $AdS_5 \times S^5$ 
background we consider, due to the presence of the magnetic field, the Hamiltonian of the giant graviton depends on the deformation parameter in a non-trivial way. 
The interplay between the conjugate momentum $J_1$ and the deformation parameter $\Delta$ creates a rich
space of parameters that we analytically examined in the previous section.

Quantum mechanically we expect the two graviton states to mix and motivated from this expectation 
we are looking for instanton solutions that  tunnel from one graviton state to the other.
In the current section we will derive explicit expressions for the instanton solutions that evolve between the
expanded three-brane state and the zero-size state. In particular, we will be mostly interested in the tunnelling from the point graviton to the giant graviton configuration since, as was discussed in the previous section, the effect of the deformation $\mu$ or equivalently $\Delta$ is to make the giant graviton the energetically favorable state. 

To begin the analysis we extend the ansatz that we introduced in \eqref{D3-ansatz} to allow for a time-dependent 
 $\rho_0$  coordinate, that is to allow for a time-dependent size of the graviton. The Lagrangian is now extended to the following expression
\begin{equation} \label{Instanton-Lagrangian}
L_{{\rm D}3}^{\rm E} = -\, \frac{N}{L^4} \, 
\left[ \rho_0^3 \, \sqrt{1 + \Gamma - \left(L^2 - \rho_0^2\right) \, 
\left(\omega^2 - \frac{\Delta^2}{L^2}\right) + \frac{\dot{\rho}_0^2}{1-\frac{\rho_0^2}{L^2}}} - \rho_0^4 \, \omega \right]
\end{equation}
where we have analytically continued to euclidean time $\tau \rightarrow i \tau$. This results in an inversion 
with respect to the $\rho_0$-axis of the potential energy of figure 1.

As a next step we perform a Legendre transformation to eliminate $\omega$ and $\Gamma$ in favor of the conserved angular momenta $J_1$ and $J_2$
and produce the Routhian
\begin{equation}  \label{Instanton-Routhian-v1}
{\cal R}^{\rm E} \, = \, \omega \, \frac{\partial L_{{\rm D}3}^{\rm E}}{\partial \omega} \, + \, \Gamma \, 
\frac{\partial L_{{\rm D}3}^{\rm E}}{\partial \Gamma} \, - \, L_{{\rm D}3}^{\rm E} \, . 
\end{equation}
Instead of working with the second order equations of motion coming from the Euclidean action \eqref{Instanton-Lagrangian}
(or the Routhian  \eqref{Instanton-Routhian-v1}), we determine the instanton solution
by the evaluation of the corresponding conserved energy
\begin{equation}  \label{Instanton-Hamiltonian-v1}
H^{\rm E} \, = \, \dot{\rho}_0 \, \frac{\partial {\cal R}^{\rm E} }{\partial \dot{\rho}_0} \, - \, {\cal R}^{\rm E} \, = \, p \, . 
\end{equation}
Following the same reasoning we have already discussed in detail in section \ref{section-Giant-Graviton}, we 
calculate from the full DBI plus WZW Lagrangian the equations of motion for the coordinates that are orthogonal to the brane worldvolume directions. 
Using the expression for the derivative of $\rho_0$ with respect to $\tau$ (i.e. $\dot{\rho}_0$) that is obtained 
from \eqref{Instanton-Hamiltonian-v1}, all the equations of motion are satisfied except from the equation of motion for $Z$.
That equation provides a constraint between the conjugate momentum $J_2$ and the conserved energy $p$, namely
\begin{equation} \label{Instanton_J2}
2 \, J_2 \, + \,p \,  \, = \, 0 \quad \Rightarrow \quad 
J_2 \, = \, \frac{N}{2 \, L} \, \frac{V(\rho_0) }{\sqrt{1 - \Delta^2 \left(1-\frac{\rho_0^2}{L^2}\right)+
\frac{\dot{\rho}_0^2}{1-\frac{\rho_0^2}{L^2}}}} \, . 
\end{equation}
In the last equation $V(\rho_0)$ is the same function defined in \eqref{Vr0}. One should, however, keep in mind that 
in \eqref{Instanton_J2} $\rho_0$ is a function of the world-volume time $\tau$ while in \eqref{Vr0} $\rho_0$ is a constant.

Substituting \eqref{Instanton_J2} in  \eqref{Instanton-Routhian-v1} we obtain the expression for the Routhian
\begin{equation}  \label{Instanton-Routhian-v2}
{\cal R}^{\rm E} \, = \, \frac{N}{L} \, \frac{V(\rho_0) }{\sqrt{1 - \Delta^2 \left(1-\frac{\rho_0^2}{L^2}\right)
+\frac{\dot{\rho}_0^2}{1-\frac{\rho_0^2}{L^2}}}} \, \left( 1+ \frac{\dot{\rho}_0^2}{1-\frac{\rho_0^2}{L^2}} \right)
\end{equation}
while substituting \eqref{Instanton_J2} in  \eqref{Instanton-Hamiltonian-v1} we obtain the expression for the Hamiltonian
\begin{equation}  \label{Instanton-Hamiltonian-v2}
H^{\rm E} \, = \, - \, \frac{N}{L} \, \frac{V(\rho_0)}{\sqrt{1 - \Delta^2 \left(1-\frac{\rho_0^2}{L^2}\right)
+\frac{\dot{\rho}_0^2}{1-\frac{\rho_0^2}{L^2}}}} \, . 
\end{equation}
Notice that setting to zero the value of the deformation parameter $\Delta$ the expressions  \eqref{Instanton-Routhian-v2}
and \eqref{Instanton-Hamiltonian-v2}, for the Routhian and the Hamiltonian respectively, flow to the undeformed 
counterparts in \cite{Grisaru:2000zn}.

\begin{figure}[ht] 
   \centering
   \includegraphics[width=7.3cm]{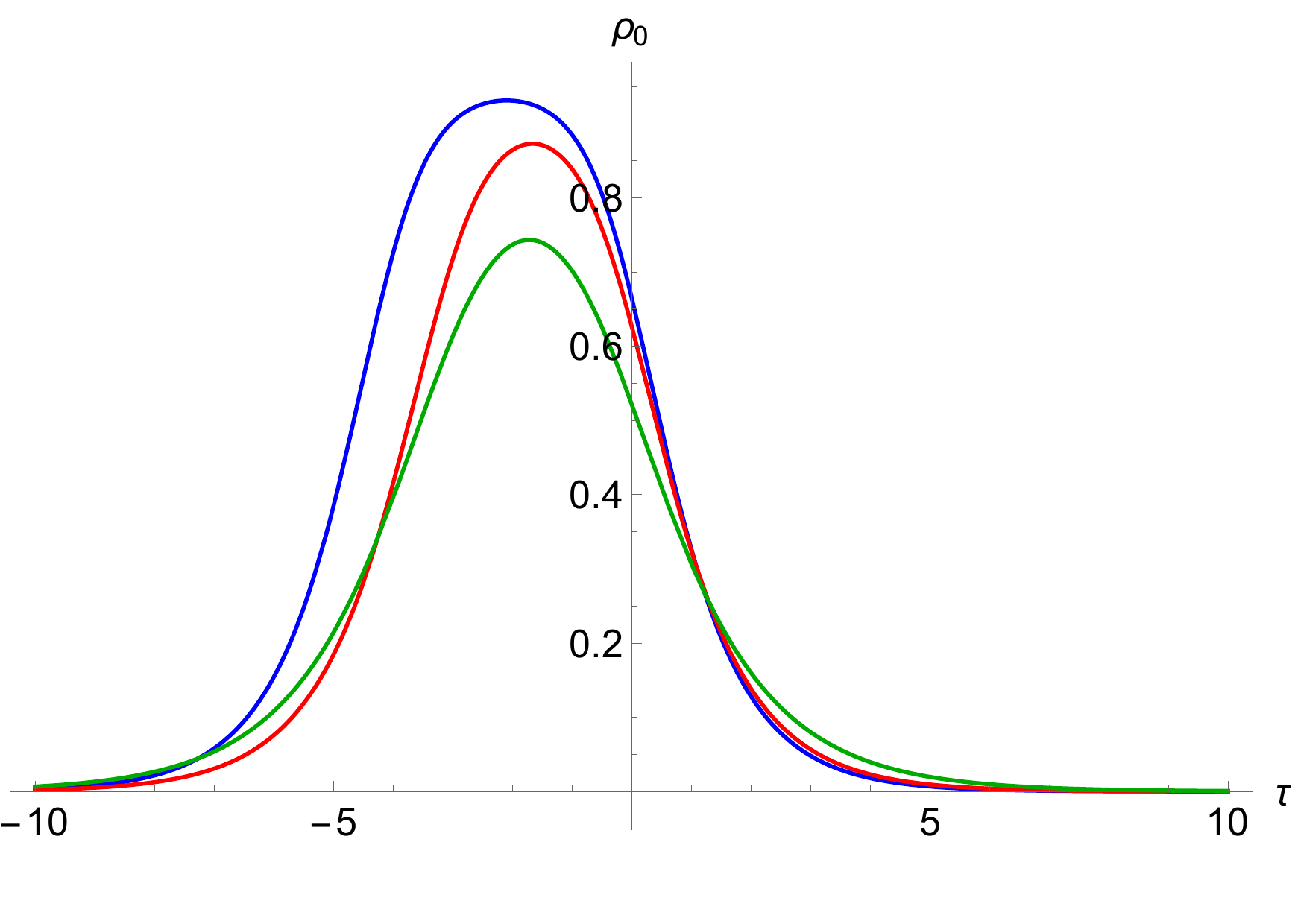}
   \hspace{0.2cm}
    \includegraphics[width=7.3cm]{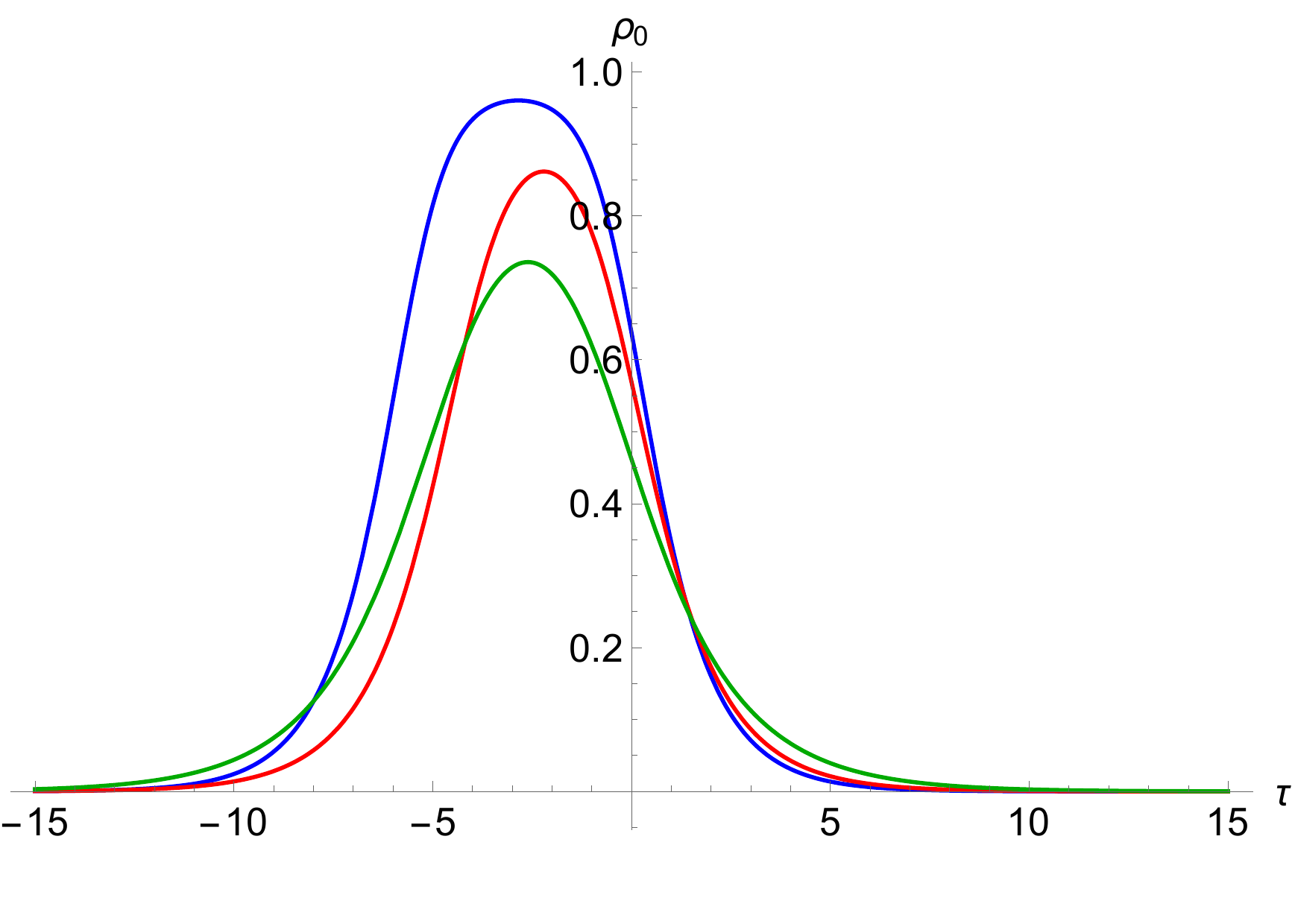}
    \caption{Plots of the instanton solution as a function of the euclidean time from equation \eqref{Instanton-solution-general}. 
    In terms of the potential that is depicted in figure  \ref{fig:Hamiltonian}, 
    the instanton transition is from the point graviton extremum at $\tau \rightarrow -\infty$ to a position of the potential 
    with the same energy as the point graviton at an intermediate time (where the maximum of the curve is located) 
    and then back to the initial point graviton extremum at $\tau \rightarrow + \infty$.
    On the left plot the conjugate momentum is fixed to the value $J_1 =0.9$ and the three curves correspond to different 
    values of the deformation parameter: Blue $\Rightarrow$ $\Delta = 0.1$, Red $\Rightarrow$ 
    $\Delta = 0.3$ \& Green $\Rightarrow$ $\Delta = 0.5$. On the right plot the conjugate momentum is fixed to the 
    value $J_1 =1.05$ and the three curves correspond to different 
    values of the deformation parameter: Blue $\Rightarrow$ $\Delta = 0.4$, Red $\Rightarrow$ 
    $\Delta = 0.5$ \& Green $\Rightarrow$ $\Delta = 0.6$. Notice that in the right plot the values of the deformation parameter 
    are always above the critical value so that the giant graviton has less energy than the point graviton. For $J_1 =1.05$ the critical value for the deformation is 
    $\Delta_{\rm crit} \approx 0.39$.}
   \label{fig:Instanton-solution-general}
\end{figure}

For the remainder of the current section we will focus on cases where the giant graviton is the energetically favored 
solution and we will construct instantons that describe tunnelling from the point to the giant graviton configuration.
In terms of the discussion we detailed in the previous section around figure \ref{fig:Hamiltonian}, there are two 
cases of interest:
\begin{itemize}
\item $J_1<1$ and $0<\Delta<1/\sqrt{2}$ (left plot of figure \ref{fig:Hamiltonian}). We remind the reader that for $\Delta>1/\sqrt{2}$ the point graviton
is not any more part of the spectrum. 
\item $J_1>1$ and $\Delta_{\rm crit}<\Delta<1/\sqrt{2}$ 
(right plot of figure \ref{fig:Hamiltonian}). 
\end{itemize}
To proceed the analysis we set $L=N=1$ and since the point graviton at $\rho_0=0$ will be the instanton starting and 
ending point, the conserved energy obtains the following value
\begin{equation} \label{conserved-energy}
H^{\rm E}(\rho_0) \, = \, H^{\rm E}(\rho_0=0) \, = \, - \, \frac{J_1}{\sqrt{\big.1 \, - \, \Delta^2}} \, . 
\end{equation}
Solving the differential equation that arises from \eqref{Instanton-Hamiltonian-v2} with the boundary condition
\eqref{conserved-energy}, we arrive at the following instanton solution
\begin{equation} \label{Instanton-solution-general}
\rho_0(\tau)  = \left[\frac{J_1 \, \big(1 -  2 \, \Delta^2\big)}{1  + e^{2 \, \left(\tau- \tau_0\right) \, \sqrt{\big.1 - 2 \, \Delta^2}} -
\Big(1 + \frac{J_1}{2}\Big)\, \Delta^2 + \frac{1}{16}\, \Delta^2 \, e^{- \, 2 \, \left(\tau- \tau_0\right)  \, \sqrt{\big.1 - 2 \, \Delta^2}} \, 
{\cal W} } \right]^{1/2}
\end{equation}
where the auxiliary quantity ${\cal W}$ is defined as follows
\begin{equation} \label{W-definition}
{\cal W} \, = \, J_1^2 \, \Delta^2\, + \, 4 \, \big(1-J_1\big) \big(1-\Delta^2\big) \, . 
\end{equation}
The constant $\tau_0$ that appears in \eqref{Instanton-solution-general} is an integration constant that gives the 
position of the instanton in euclidean time, it is in fact the zero mode that determines the instanton position.
In the rest of the analysis and for all the plots we fix this constant to zero. i.e. 
$\tau_0=0$. In figure \ref{fig:Instanton-solution-general} we plot the instanton solution for two values of the conjugate 
momentum $J_1$, one less than one and one larger than one, and three different values of the deformation parameter $\Delta$. 

\begin{figure}[ht] 
   \centering
   \includegraphics[width=7.3cm]{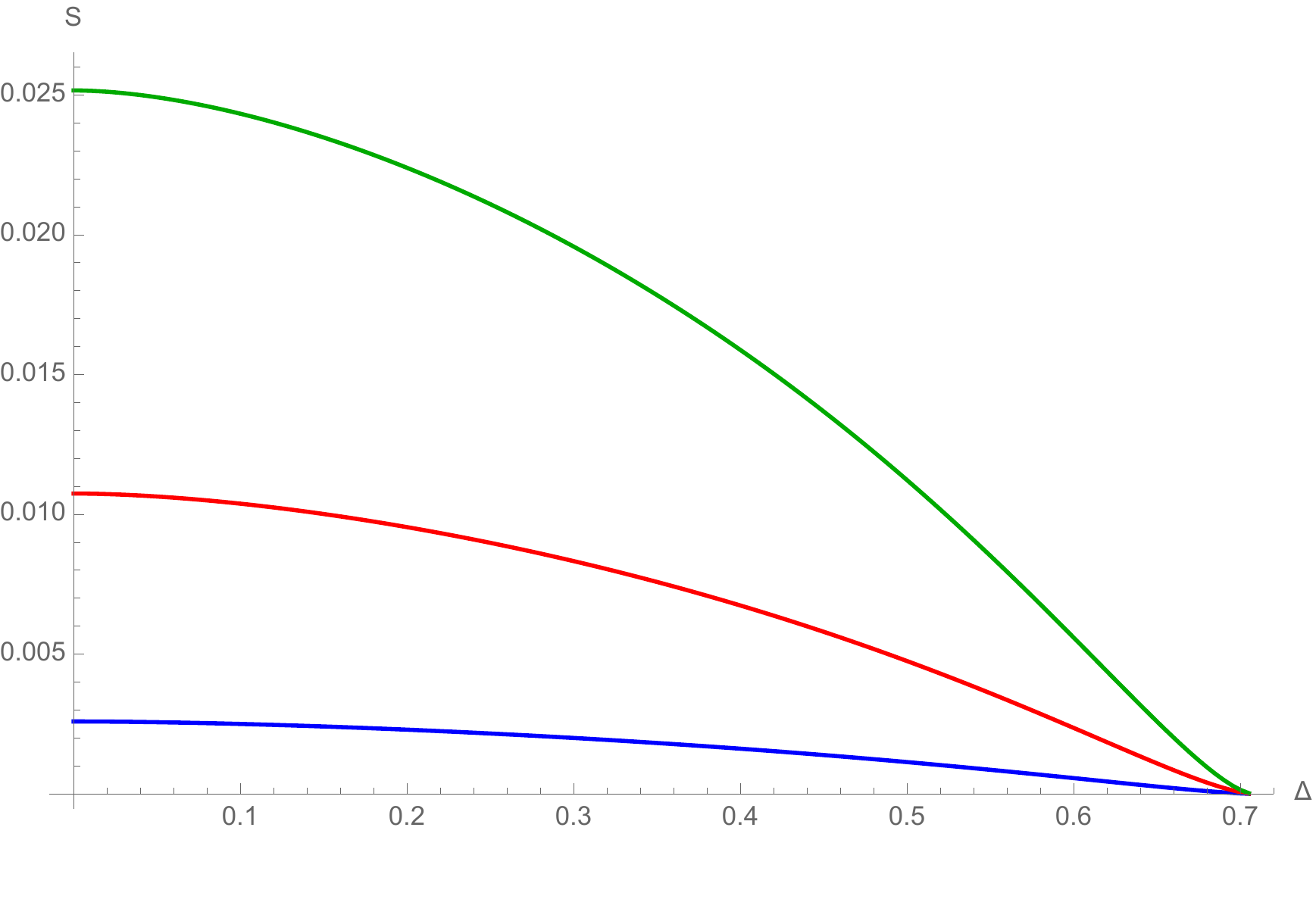}
   \hspace{0.2cm}
    \includegraphics[width=7.3cm]{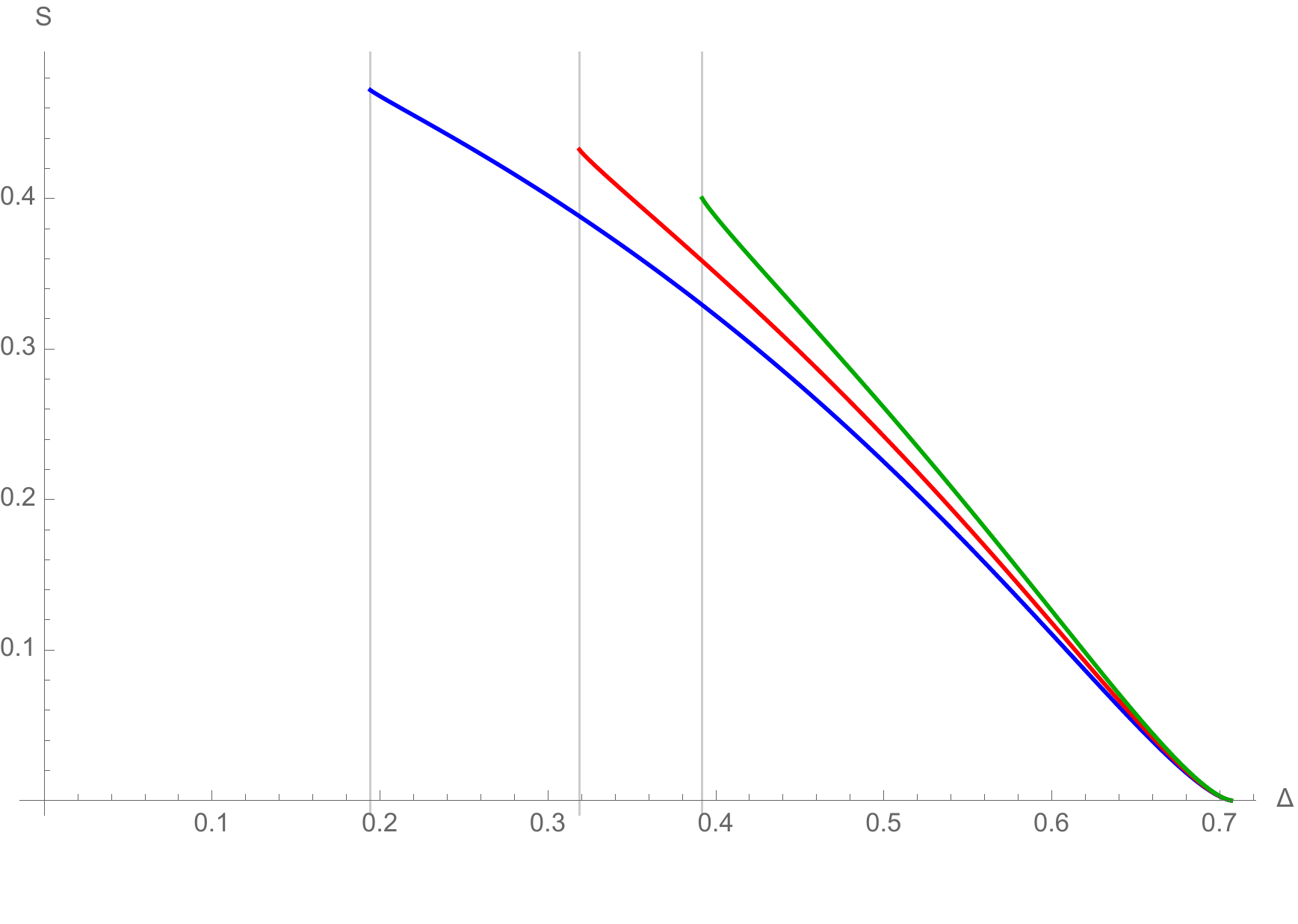}
    \caption{Total action as a function of the deformation parameter $\Delta$ for values of $J_1<1$ (left plot) and for values $J_1> 1$ (right plot). 
    When $J_1<1$, $\Delta$ takes all the values between 0 and $1/\sqrt{2}$, while for $J_1> 1$ $\Delta$ is restricted to the 
    values that are above the red line in figure \ref{DeltavsJ}. The gridlines emphasize  those values, that are the lower limits 
    for $\Delta$ given a value for $J_1$. 
    The correspondence between colors and values of $J_1$ for the left plot is: 
    Blue $\Rightarrow$ $J_1 = 0.1$, Red $\Rightarrow$ 
    $J_1 = 0.2$ \& Green $\Rightarrow$ $J_1 = 0.3$. 
    The correspondence between colors and values of $J_1$ for the right plot is: Blue $\Rightarrow$ 
    $J_1 = 1.01$, Red $\Rightarrow$ 
    $J_1 = 1.03$ \& Green $\Rightarrow$ $J_1 = 1.05$. }
   \label{fig:ActionVSJ1}
\end{figure}

\begin{figure}[ht] 
   \centering
   \includegraphics[width=8cm]{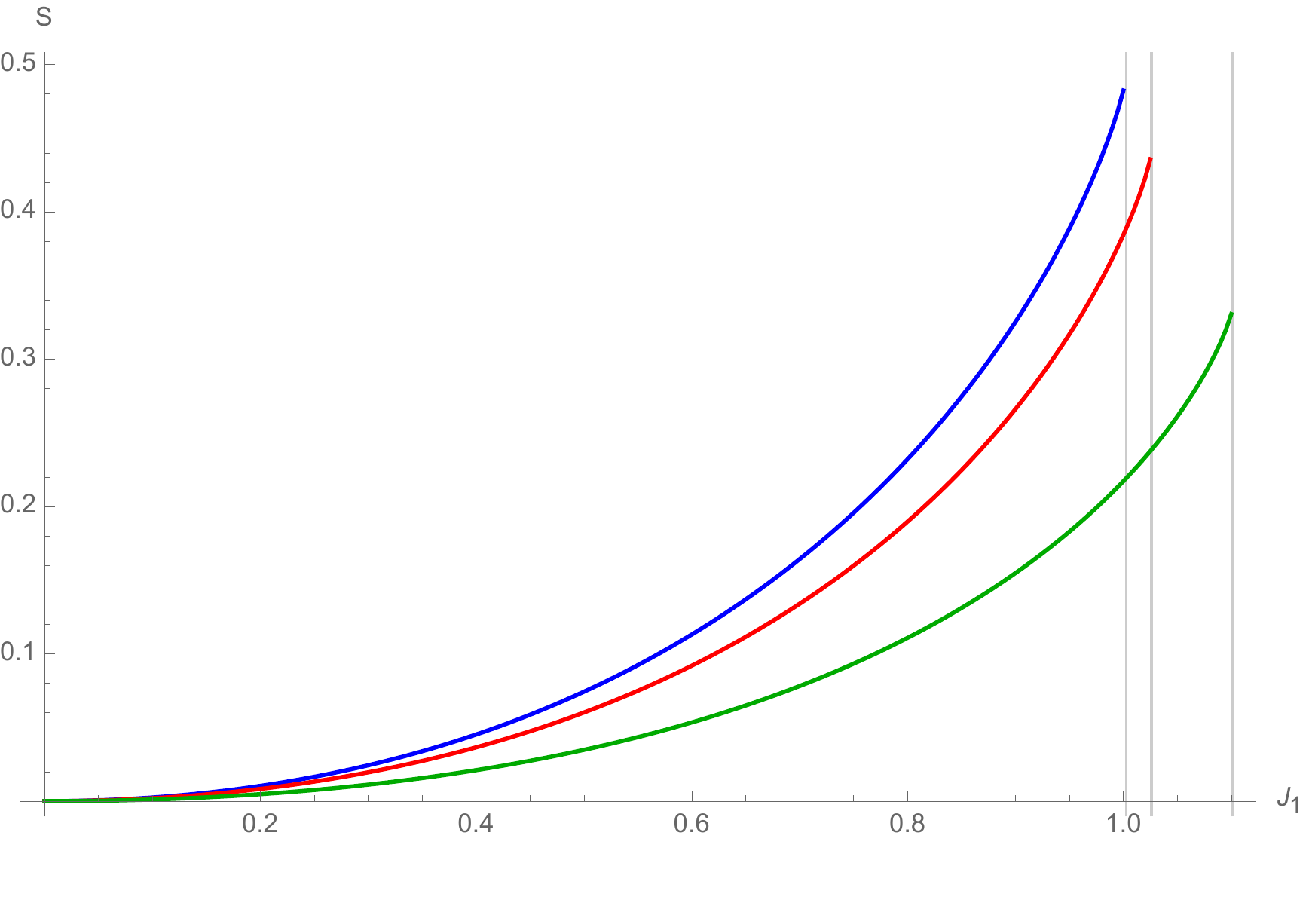}
    \caption{Total action as a function of $J_1$ for different values of $\Delta$. The upper bound for the value of $J_1$ depends
    on $\Delta$, as it is depicted in figure \ref{DeltavsJ}. $J_1$ should be always above the red line
    and the gridlines emphasize  the maximum values for $J_1$. 
    The correspondence between colors and values of $\Delta$ is:
    Blue $\Rightarrow$ $\Delta = 0.1$, Red $\Rightarrow$ $\Delta = 0.3$ \& Green $\Rightarrow$ $\Delta = 0.5$.}
    \label{fig:ActionVSDelta}
\end{figure}

The next step is the calculation of the euclidean action. Since the Routhian ${\cal R}^{\rm E}$ 
\eqref{Instanton-Routhian-v2}  evaluated around the point graviton position remains finite  the 
evaluation of the integral $\int d\tau {\cal R}^{\rm E}$ diverges. To regularize the integral we subtract the Routhian 
at the point graviton extremum and the relevant quantity to calculate is the following 
\begin{equation} \label{Instanton-TotalAction-general}
{\cal S}\, = \, \int_{-\infty}^{\tau_{\rm turn}} d\tau \, \Delta {\cal R} 
\quad \text{with} \quad
\Delta {\cal R} \, = \, {\cal R} (\rho_0) \, - \, {\cal R} (\rho_0 = 0 )
\end{equation}
where $\tau_{\rm turn}$ is the time that corresponds to the maximum of the curve that is depicted in figure 
\ref{fig:Instanton-solution-general}. Notice that the upper limit of integration is the turning point $\tau_{\rm turn}$ since if the graviton manages to tunnel up to this point then it will immediately roll to the global minimum of the potential which is the giant graviton state.
In figures \ref{fig:ActionVSJ1} and \ref{fig:ActionVSDelta} we have plotted the 
total action (only numerical evaluation of the integral in \eqref{Instanton-TotalAction-general} is possible) 
as a function of the conjugate momentum $J_1$ and the deformation parameter $\Delta$. Calculation of the 
total action provides a measure for the degree of tunnelling. More precisely, the transition probability from the point graviton to the stable giant graviton is proportional to the exponential of minus the action of \eqref{Instanton-TotalAction-general}.
Two comments are in order. Firstly, notice that the transition probability is exponentially small since if we reinstate the number of colours $N$, which has been set to unity in the numerical calculation above, one gets 
$P\sim e^{-N s}$, where $s>0$. Secondly, note that figure 4 implies that as the deformation parameter increases the transition probability increases since $\cal S$ becomes smaller. 
This is fully consistent with the shape of the potential depicted in figure \ref{fig:Hamiltonian}. 
It is evident from that figure that as the deformation parameter increases the height
of the barrier which the brane has to tunnel through becomes smaller making the tunnelling more probable.

\subsection{Special Instanton solution}

\begin{figure}[ht] 
   \centering
   \includegraphics[width=8cm]{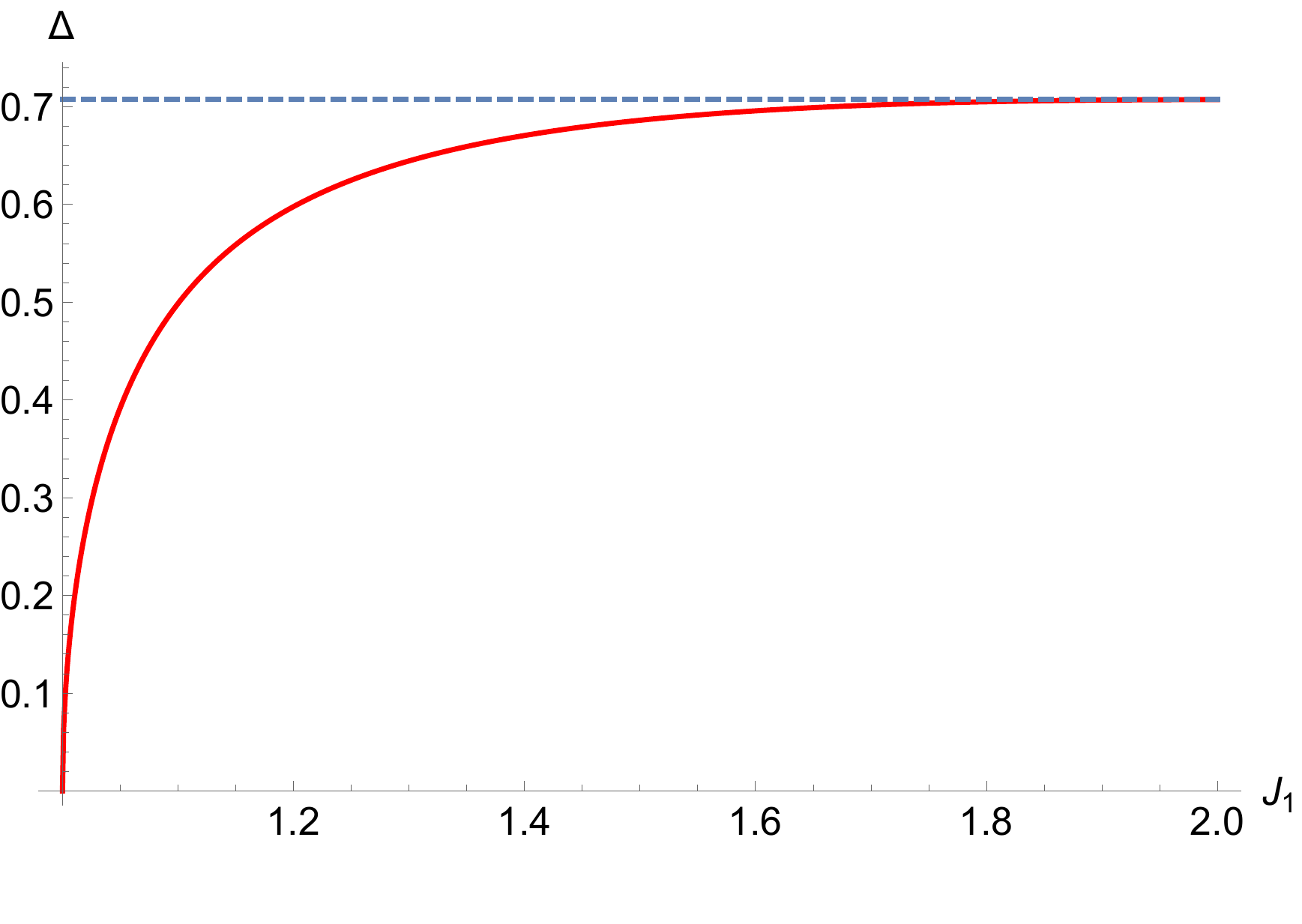}
    \caption{Plot of the relation between $\Delta$ and $J_1$ as depicted in equation \eqref{DvsJ_equation}. The tunnelling 
    from the point to the giant graviton occurs only when the pair ($J_1$, $\Delta$) is above the red line. Below the red line the point graviton is energetically favoured and the tunnelling should be in the opposite direction.}
   \label{DeltavsJ}
\end{figure}

In this subsection, we will examine a  special class of instanton solutions among those that are described by
the analytic expression that is presented in \eqref{Instanton-solution-general}. Setting to zero the value of the 
auxiliary parameter $\cal W$ that is defined in \eqref{W-definition} introduces a constraint between the 
conjugate momentum $J_1$ and the deformation parameter $\Delta$. Solving the constraint with respect to $\Delta$
means that by choosing the value of $J_1$ we automatically fix the value of the deformation parameter $\Delta$ as follows
\begin{equation} \label{DvsJ_equation}
{\cal W} \, = \, 0  \quad \Rightarrow \quad \Delta \, = \, \Bigg[1 \, - \, \frac{1}{4}\, \frac{J_1^2}{1 \, - \, J_1}\Bigg]^{-1/2} \, . 
\end{equation}
In figure \ref{DeltavsJ} we plot this relation between $\Delta$ and $J_1$.

\begin{figure}[ht] 
   \centering
   \includegraphics[width=7.3cm]{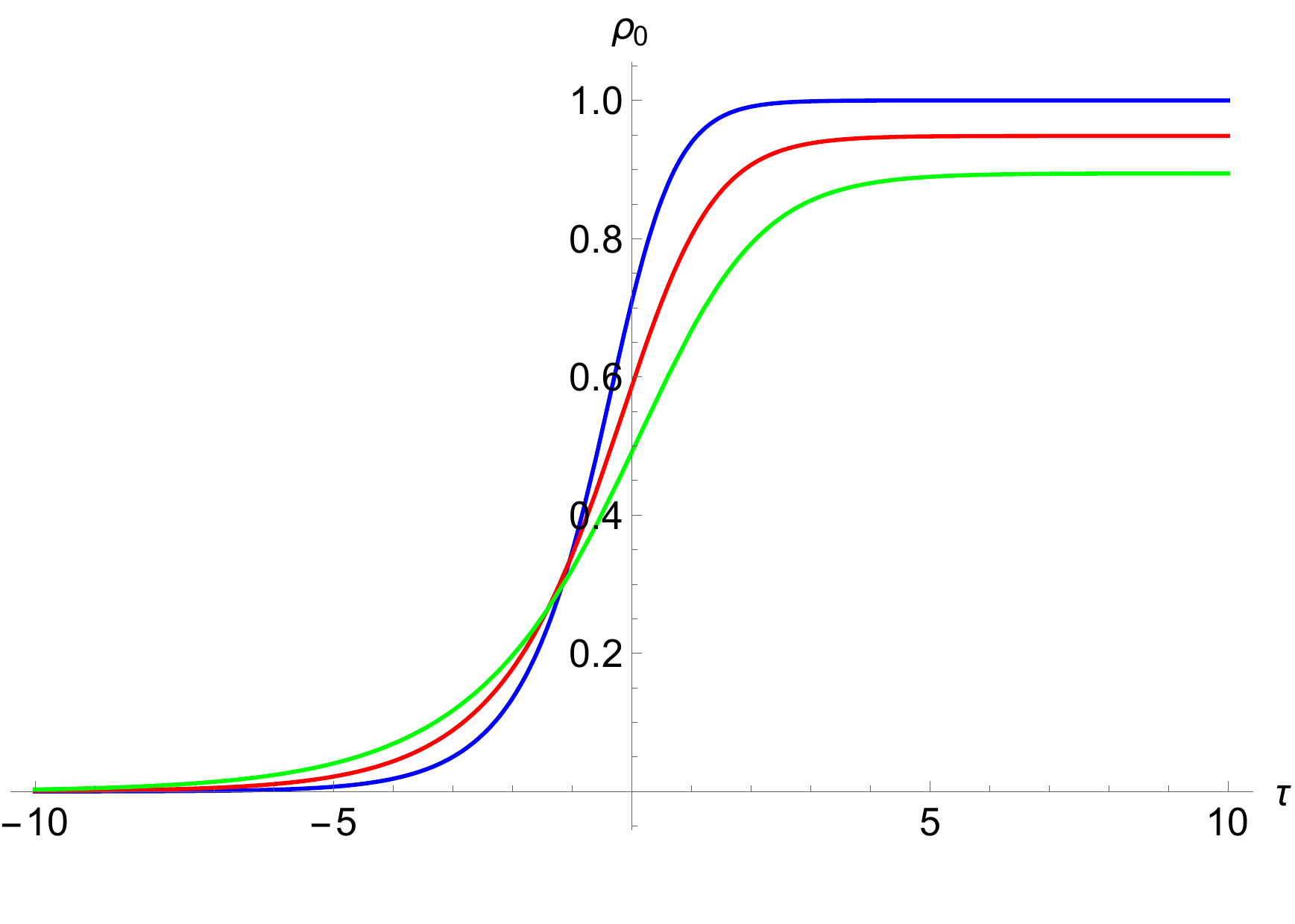}
   \hspace{0.2cm}
    \includegraphics[width=7.3cm]{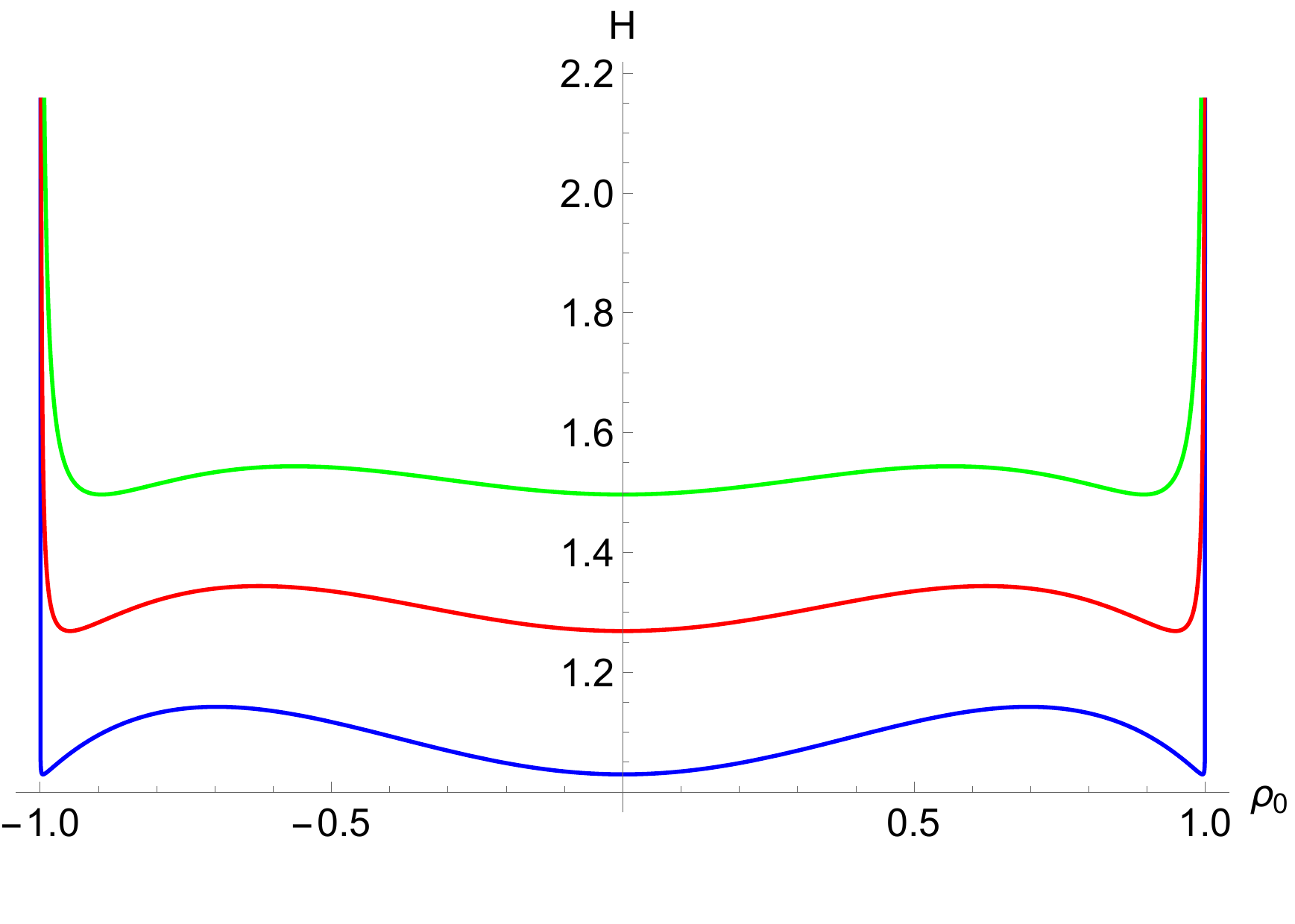}
    \caption{Left plot: Instanton solution in the specific case that ${\cal W} =0$ in which the two graviton solutions are degenerate. The plot is for 3 different values of $J_1$. 
    Right plot: Hamiltonian for the instanton 
    solutions of the left plot. It is clear that when ${\cal W} =0$ the energies of the point and giant gravitons are equal. 
    The correspondence between colors and values of $J_1$ is:
    Blue $\Rightarrow$ $J_1= 1.01$, Red $\Rightarrow$ $J_1 = 1.1$ \& Green $\Rightarrow$ $J_1 = 1.2$}
   \label{fig:Special-Instanton-solution-Energy}
\end{figure}

Implementing the constraint \eqref{DvsJ_equation} on the general instanton solution of  \eqref{Instanton-solution-general}
and calculating the limits at $\tau \rightarrow - \, \infty $ and $\tau \rightarrow + \, \infty $ we obtain the following
results
\begin{equation}
\rho_0 \rightarrow 0 \quad \text{for} \quad \tau \rightarrow - \, \infty 
\quad \& \quad 
\rho_0 \rightarrow \sqrt{\big.2 - J_1} \quad \text{for} \quad \tau \rightarrow + \, \infty \, .
\end{equation}
The last two relations fully determines the range of values for both $J_1$ and $\Delta$. More specifically, 
we have that $\Delta \in [0, 1/ \sqrt{2})$ and $J_1 \in [1, 2)$. On the left plot of figure \ref{fig:Special-Instanton-solution-Energy} 
the special instanton solution for ${\cal W} =0$ is presented. It corresponds to an instanton transition from the zero-size
point graviton solution at $ \tau \rightarrow - \, \infty$ to the expanded giant graviton solution at $ \tau \rightarrow + \, \infty$.
Notice that another instanton solution, symmetric with respect to the vertical axis at $ \tau = 0$, exists that would correspond 
to a transition from the expanded giant graviton solution
at $ \tau \rightarrow - \, \infty$ to the zero size point graviton solution at $ \tau \rightarrow + \, \infty$. 
From the moment that eventually, when increasing the deformation parameter, 
the energetically favored solution is the giant graviton, we are not interested in 
transitions from the giant to the point graviton solution.  

On the right plot of figure \ref{fig:Special-Instanton-solution-Energy} we plot the Hamiltonian of the special instanton 
solution. In that way a remarkable feature of the special solution is uncovered. When the constraint  
\eqref{DvsJ_equation} is satisfied, the energies of the point graviton and of the giant graviton are equal.  
This situation is analogous to the instanton calculation in the undeformed case of ${\cal N}=4$ \cite{Grisaru:2000zn}, 
where the energies of the point and the giant graviton were also equal. Here, tuning the relation between $\Delta$ 
and $J_1$, we are able to identify similar situations where the two solutions have the same energy. The important difference 
is that increasing the deformation parameter will eventually lead to an energetically favored giant graviton.

\begin{figure}[ht] 
   \centering
   \includegraphics[width=8cm]{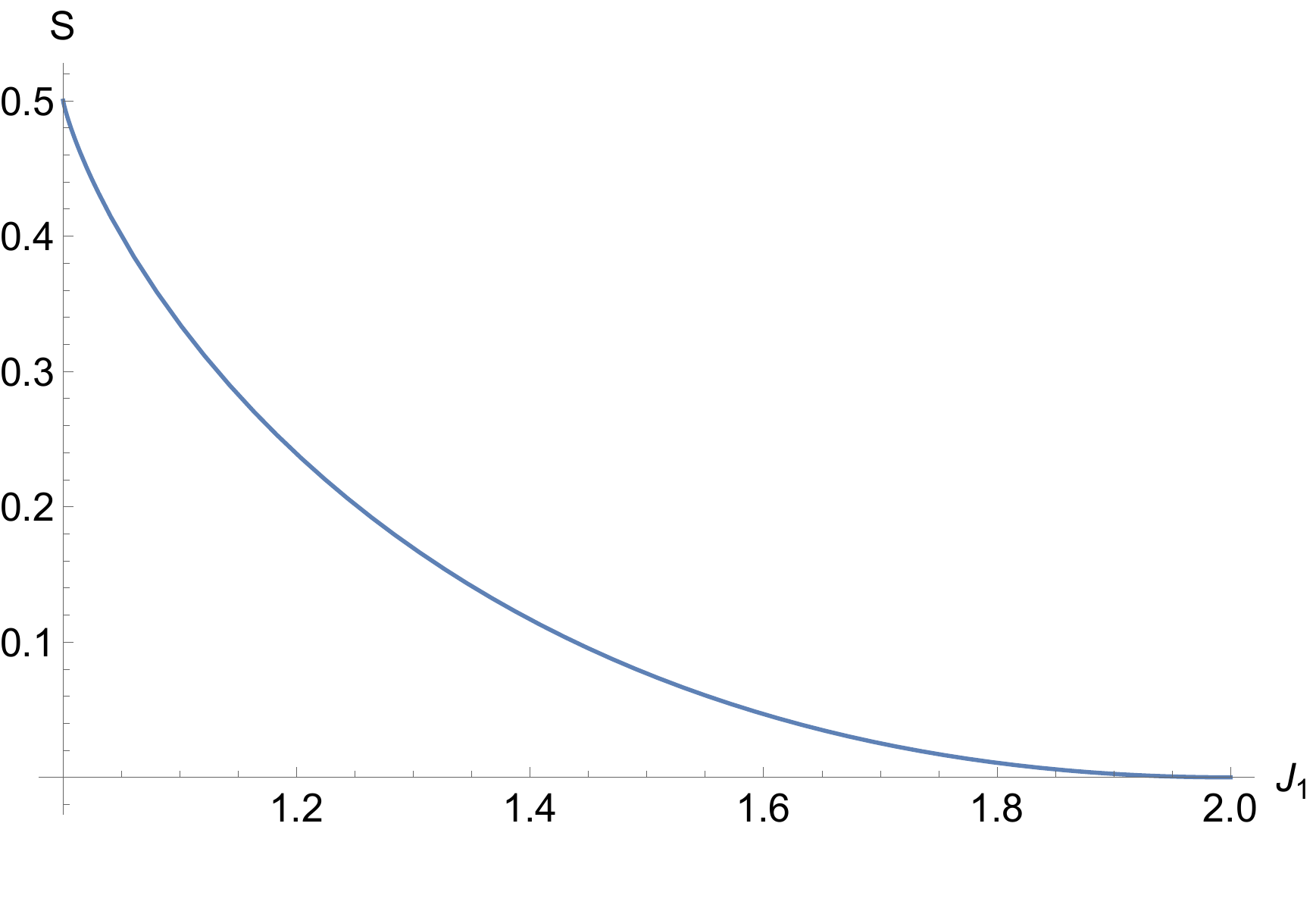}
    \caption{Total action as a function of $J_1$, in the special case of ${\cal W} =0$ from equation 
      \eqref{special-instanton-TotalAction}. $J_1$ and $\Delta$ are related 
    through equation \eqref{DvsJ_equation}.}
    \label{fig:ActionVSJ-special-instanton}
\end{figure}

The next important step is to calculate the total action of the special solution. Regularizing the Routhian as before 
in \eqref{Instanton-TotalAction-general} but now integrating from $-\infty$ to $+\infty$ we are able to obtain an 
analytic result for the total action
\begin{eqnarray}  \label{special-instanton-TotalAction}
{\cal S}\, = \, \int_{-\infty}^{+ \infty} d\tau \, \Delta {\cal R} & = & \int_0^{\sqrt{2-J_1}} d\rho \, 
\frac{\rho \, \big(2 - \rho^2 -  J_1\big)}{1 - \rho^2}  
\nonumber \\
& = & \frac{1}{2}\, \Bigg[ 2 - J_1 + \big(J_1 - 1\big) \, \log \big(J_1 - 1\big)\Bigg] \, . 
\end{eqnarray}
In figure \ref{fig:ActionVSJ-special-instanton} we plot the total action as a function of the angular momentum $J_1$, from where it is 
evident that it monotonically decreases. Thus, as $J_1$ ( or $J_1/N$, if we reinstate the number of colours) increases, the action decreases 
and the probability of tunnelling is increasing. This qualitative picture is also supported by the plot of the Hamiltonian (see the right plot of figure 7), where 
one can see that when $J_1$ increases, the difference between the extremum (minimum) of the point graviton and the 
nearby extremum (maximum), gets smaller. The potential becomes more shallow and as a result the probability
of tunnelling becomes larger.

\section{Three-point function of two giant gravitons and a dilaton mode}\label{section:3-point-function}

In this section, we will focus on the  holographic calculation of the three-point correlation function involving two operators dual to the giant graviton and one operator dual to one of the dilaton modes.\footnote{The method to calculate three-point correlators of the heavy-heavy-light type holographically, as well as several important results can be found in \cite{Zarembo:2010rr,Costa:2010rz,Georgiou:2010an,Georgiou:2011qk,Georgiou:2013ff,Bajnok:2016xxu}.} 
The result will  provide the strong coupling prediction for the aforementioned three-point correlator. 
Similar holographic calculations involving giant magnons or spiky strings of the Schrodinger background have been 
performed recently in \cite{Georgiou:2018zkt}. In addition, the holographic computation of heavy-heavy-light correlation 
functions with the heavy operators being the giant gravitons and the light operators being BPS states
of the undeformed $AdS_5\times S^5$ background have been presented in \cite{Bissi:2011dc}.

In order to proceed with the holographic calculation we need a giant graviton solution that tunnels from one point of 
the boundary (this is one of the points where the dual field theory operator is situated) to another.  
To this end we construct a slight generalization of the giant graviton solution presented in section 
\ref{section-Giant-Graviton}. This reads
\begin{eqnarray}\label{gen-sol}
&& 
T\,=\, \kappa \,\tau \, , 
\quad
V \,=\,\nu \, \tau \,  - \, \frac{\vec X^2_0}{4} \, \sin 2 \, \kappa \, \tau \, , 
\quad 
{\vec X} \,=\, {\vec X}_0 \, \sin \kappa \, \tau \, ,
\nonumber \\ [5pt]
&&
Z \,=\, Z_0 \, , 
\quad 
\rho = \frac{\rho_0}{L}
\quad \& \quad
\chi= \omega \, \tau,
\end{eqnarray}
where the constants appearing in \eqref{gen-sol} are given by
\begin{eqnarray}\label{constants}
&& \kappa = \frac{1}{L} \, , \quad 
\nu = \frac{Z_0^2}{L\, {\cal Q}} \, \left(\frac{J_1}{N}-1\right)^2 \, , \quad 
\omega \, = \, \frac{2}{L \, \sqrt{\cal Q}} \, \left(1- \frac{J_1}{4 \, N} \right) 
\nonumber \\ [5pt]
&&
\frac{\rho_0^2}{L^2} \, = \, 2- \frac{J_1}{N}\, , \quad 
\mu^2 \, = \, \frac{{Z_0^4}}{{\cal Q}} \, \left(\frac{J_1}{N}-1 \right)
\quad {\rm with} \quad 
{\cal Q} = \frac{J_1^2}{4 \, N^2}\,+\, \frac{J_1}{N}-1.
\end{eqnarray} 
Notice that we have turned on two additional coordinates of the Schrodinger part of the spacetime, namely $\vec X$. 
As can be seen from \eqref{gen-sol} this induces a modification of the $V$ coordinate, too.\footnote{Let us also mention that the three-point function calculation of this section can be performed analytically only for the special case where the deformation parameter $\Delta$ and the angular momentum $J_1$ are related through \eqref{DvsJ_equation}, that is when the point and the giant graviton are degenerate.  The most general case can be treated in a similar way but the analogous to \eqref{constants} expressions become very lengthy without gaining no further insight.}

The next step is to use the following coordinate transformation in order to pass from the global to the Poincare coordinates 
\cite{Blau:2009gd}
\begin{equation} \label{transf}
x^+ =\tan T \, , \qquad 
x^-=V-\frac{1}{2} \left(Z^2+\vec{X}^2\right) \, \tan T \, , \qquad 
z=\frac{Z}{\cos T} \, ,\qquad 
\vec{x}=\frac{\vec{X}}{\cos T}
\end{equation}
and subsequently to perform an analytic continuation to the world-sheet time $\tau \rightarrow i\, \tau$ 
after which the solution takes the form
\begin{eqnarray} \label{an-sol}
&&
\mathbf{x} \, = \, \frac{\mathbf{X}}{2}\,\tanh\kappa \tau \, , 
\hspace{1cm}
x^+ \, = \, i \, \frac{\mathbf T}{2}\,\tanh\kappa \tau \, , 
\hspace{1cm}
z \, = \, \sqrt{\frac{\mathbf T}{2}}\, \frac{Z_0}{\cosh\kappa \tau }\, ,
\nonumber \\[5pt]
&& x^- \, = \, - \, \frac{i}{2} \, \left[Z_0^2 \, - \, 
\frac{\mathbf{X}^2}{2\, \mathbf T}\right]\,\tanh\kappa \tau \, + \, i \, \nu \, \tau  \, , 
\quad 
\rho = \frac{\rho_0}{L}
\quad \& \quad
\chi \,= \, i \, \omega \, \tau.
\end{eqnarray}
This is the appropriate giant graviton solution travelling from one point of the boundary at $\tau=-\infty$ 
to another point at $\tau=+\infty$.
In what follows, we will also need the on-shell value of the D3 brane Lagrangian which can be evaluated from \eqref{D3-action}.
It reads
\begin{equation} \label{on-shell}
{\cal L}_{\text{on shell}}^{\text{GG}} \, =  \, i \, \frac{N}{4 \, \pi^2 \, L \, \sqrt{\cal Q}}  \, 
\left(1- \frac{J_1}{2 \, N}\right)^2 \, \sin \theta \, . 
\end{equation}

One may now define the the following kinematic Schrodinger invariant
\begin{equation} \label{v12}
v_{12}
=-{1 \over 2}
\left(\frac{x^2_{12}}{t_{12}}+\frac{x^2_{23}}{t_{23}}-\frac{x^2_{13}}{t_{13}} \right),
\end{equation}
where $t_{ij} = t_i-t_j$ and similarly $\vec{x}_{ij}=(\vec{x}_i-\vec{x}_j)$.
In \eqref{v12}, $(\vec x_1,t_1)$ and $(\vec x_2,t_2)$ denote the positions of the heavy operator on the boundary 
while $(\vec x_3,t_3)$ the position of the dilaton mode. In what follows and in order to keep the expressions for the 
three-point correlator as simple as possible we take the time ordering of the operators to be as follows
\begin{equation} \label{t3infty}
t_3 < t_1 < t_2 \quad \& \quad t_3 \rightarrow - \, \infty \, .
\end{equation}
As a result the invariant cross ratio simplifies to  $v_{12}=-{1 \over 2}
\frac{x^2_{12}}{t_{12}}={1 \over 2}
\frac{\mathbf X^2}{\mathbf T}$. One may use this last relation to rewrite, by taking into account that
the expression for $Z_0$ becomes
\begin{equation}
Z_0^2 \, = \, \frac{v_{12}}{2} \, \Bigg[1- \frac{E}{2\,N \, \sqrt{\cal Q}}\Bigg]^{-1} \, , 
\end{equation}
the analytically continued giant graviton solution \eqref{an-sol} in terms of the invariant $v_{12}$, $\mathbf{X}$ and $\mathbf T=t_2-t_1$.
It reads
\begin{eqnarray} \label{an-sol-1}
&&
\mathbf{x} \, = \, \frac{\mathbf{X}}{2}\,\tanh\kappa \tau \, , 
\hspace{1cm}
x^+ \, = \, i \, \frac{\mathbf T}{2}\,\tanh\kappa \tau \, , 
\hspace{1cm}
z \, = \,\frac{\sqrt{{\mathbf T}\, v_{12}}}{2 \, \cosh\kappa \tau }\,
\Bigg[1- \frac{E}{2\,N \, \sqrt{\cal Q}} \Bigg]^{-1/2}\, ,
\nonumber \\[5pt]
&& x^- \, = \, i \, \nu \, \tau \, + \, \frac{i \, v_{12}}{2} \, \left[1 \, - \,\frac{1}{2} \,
\Bigg[1- \frac{E}{2\,N \, \sqrt{\cal Q}}\Bigg]^{-1}
\right]\,\tanh\kappa \tau  \, , 
\quad 
\rho = \frac{\rho_0}{L} \, , 
\quad
\chi \,= \, i \, \omega \, \tau.
\end{eqnarray}

The last piece of information needed is the bulk-to-boundary propagator of the light field, that is of the dilaton mode. 
In the limit \eqref{t3infty} it takes the form \cite{Georgiou:2018zkt}
\begin{equation} \label{propagator}
K(z, \vec{x},t; \vec{x}_3, t_3) \approx \frac{i\left( \frac{\mu \, M_3}{2}\right)^{\Delta_3-1} 
e^{-{i \over 2}\pi \Delta_3}}{\pi \, \Gamma(\Delta_3-2)} 
\left(\frac{z}{- \, t_3}\right)^{\Delta_3}\, ,
\end{equation}
where $M_3$ denotes the mass eigenvalue of the dilaton mode participating in the three-point correlator.

Now, we have all the ingredients in order to calculate the three-point correlator.
As discussed, in some detail, in a similar calculation
\cite{Georgiou:2018zkt} the ratio of the three-point to the two-point correlation functions is given by
\begin{equation} \label{recipe}
\frac{G_{3}(\bar{x}_1, \bar{x}_2, \bar{x}_3)}{G_{2}(\bar{x}_1, \bar{x}_2)} \,  =-
\int d^4 \xi \, e^{i \, M_3 \, x^{-}\left(\tau, \sigma\right)} \, K(\bar{x}_{\rm classical}(\tau,\sigma); \bar{x}_3) \,
 {\cal L}_{\text{on shell}}^{\rm classical}\,.
\end{equation}
Notice that the result for the three-point function takes the form of a vertex operator that corresponds to the 
light state integrated over the classical world-volume of the D3-brane which represents the giant graviton at the strong 
coupling regime.\footnote{The minus sign in front of the right hand side of \eqref{recipe} originates from the 
functional differentiation of the brane action with respect to the dilaton field (see \eqref{D3-action}).}

Plugging \eqref{on-shell} and \eqref{propagator} and using the expressions for
the solution \eqref{an-sol-1} in \eqref{recipe} 
one may perform the integration over the three-sphere along which the D3-brane extends to obtain
\begin{equation}
\frac{G_{3}}{G_{2}} =  \, \frac{N \, v_{12}^{\frac{\Delta_3}{2}}}{\sqrt{\cal Q}}\,
\frac{\left[1- \frac{J_1}{2 \, N}\right]^2}{\Big[1- \frac{E}{2\,N \, \sqrt{\cal Q}}\Big]^{\frac{\Delta_3}{2}} } \, 
\frac{\left( \frac{M_3}{2}\right)^{\Delta_3-1}e^{-{i \over 2}\pi \Delta_3}}{2^{\Delta_3- 2} \, 
\pi \, \Gamma(\Delta_3-2)} \, 
\left(\frac{\sqrt{t_{21}}}{-\, t_3}\right)^{\Delta_3}\frac{1}{L} \, \int^{+\infty}_{-\infty} \frac{e^{i \, M_3 \, x^{-}\left(\tau, \sigma\right)}}
{\left(\cosh \frac{1}{L}\, \tau \right)^{\Delta_3} } \, d\tau \, .
\end{equation}
The remaining $\tau$ integral can be approximated by
\begin{equation}
{\cal I} \, \approx \, \frac{1}{L} \, \int^{+\infty}_{-\infty} \frac{e^{\frac{\alpha}{L} \, \tau}}
{\left(\cosh \frac{1}{L}\, \tau \right)^{\Delta_3} } \, d\tau\, 
\quad {\rm where} \quad 
\alpha = - \frac{v_{12} \, M_3}{2} \, \Bigg[1 + \frac{{\cal Q} - 2 \left(\frac{J_1}{N} -1\right)^2}{\frac{E}{N} \, 
\sqrt{\cal Q} -2 \, {\cal Q}}\Bigg] \, . 
\end{equation}
The approximation is done in the exponential $e^{i \, M_3 \, x^{-}\left(\tau, \sigma\right)}$ of \eqref{recipe} and consists of replacing the 
$\tanh{(\tau/L)}\approx \tau/L$ appearing in $x^{-}$. This is a legitimate approximation since the bulk-to-boundary propagator in \eqref{recipe} has a peak around $\tau=0$ and falls off exponentially away from this value.
As a result the  integral can be evaluated to give
\begin{equation}
{\cal I} \, = \, 2^{\Delta_3 -1} \,
\mathrm{B} \left[\frac{1}{2} \left(\Delta_3 - \alpha \right), \frac{1}{2} \left(\Delta_3 +\alpha \right)\right] \, . 
\end{equation}

As discussed in \cite{Georgiou:2018zkt} (see eq. (3.9)) the ratio of the three to the two-point correlator takes in the limit \eqref{t3infty} the following form 
\begin{equation}\label{3/2}
\frac{G_{3}}{G_{2}} = \left(\frac{\sqrt{t_{21}}}{-\, t_3}\right)^{\Delta_3}\ \, \tilde F(v_{12})\, ,
\end{equation}
where $\tilde F(v_{12})$ is an undetermined function of the Schr\"{o}dinger invariant $v_{12}$ that can not be determined by conformal invariance. It reads
\begin{equation}
\tilde F(v_{12})=  \, \frac{N \, v_{12}^{\frac{\Delta_3}{2}}}{\sqrt{\cal Q}}\,
\frac{\left[1- \frac{J_1}{2 \, N}\right]^2}{\Big[1- \frac{E}{2\,N \, \sqrt{\cal Q}}\Big]^{\frac{\Delta_3}{2}} } \, 
\frac{\left( \frac{M_3}{2}\right)^{\Delta_3-1}e^{-{i \over 2}\pi \Delta_3}}{2^{-1} \, 
\pi \, \Gamma(\Delta_3-2)} \, 
\,
\mathrm{B} \left[\frac{1}{2} \left(\Delta_3 - \alpha \right), \frac{1}{2} \left(\Delta_3 +\alpha \right)\right]\, .
\end{equation}

Our calculation provides the strong coupling prediction for the three-point correlation function and is fully consistent with the three-dimensional conformal symmetry of the Schr\"{o}dinger group of the theory.

\section{Conclusions}\label{section:conclusions}

The center of attention of the current paper is the construction and further study of a giant graviton configuration in the framework of the Schr\"{o}dinger holography. 
On the gravity side the giant gravitons are described by a D3-branes that extend along the time direction of the $Sch_5$ part of the geometry and 
wrap a three-sphere inside the internal space of the undeformed five-sphere. The main result of this paper is to confirm, in the original Schr\"{o}dinger spacetime, the picture discovered previously in \cite{Georgiou:2020qnh} for the case of the pp-wave limit  of the Schr\"{o}dinger geometry , namely that 
the giant graviton configuration becomes the energetically favored stable configuration compared to the point graviton one. This fact is remarkable by itself since in most of the literature the tendency is exactly the opposite, the point graviton is the ground state of the D3-brane. The aforementioned fact leads also to the possibility of tunnelling from the point to the giant graviton configuration. We calculate, explicitly, the instanton  solution and its corresponding action which gives a measure of the tunnelling probability. Finally, we evaluate holographically the 3-point correlation function of two giant gravitons and one dilaton mode.

The giant graviton solution depends on the deformation parameter of the spacetime $\Delta$ and on the angular momentum $J_1$ along the sphere equator.
We vary the value of the deformation parameter keeping the angular momentum fixed, either below or above 
the constant value $J_1=1$. In the former case and for non-zero values of the deformation parameter the degeneracy between the 
point and the giant graviton, that was present in the undeformed ${\cal N}=4$ case \cite{Grisaru:2000zn}, 
is lifted in favor of the giant graviton. Moreover, there is a maximum value of the deformation parameter 
(e.g. $\Delta = 1/ \sqrt{2}$) above which the point 
graviton solution ceases to be at a minimum of the potential and evetually
disappears from the spectrum. Notice that both aforementioned features were present also in the analysis of 
the giant graviton solution in the pp-wave limit of the  Schr\"{o}dinger background \cite{Georgiou:2020qnh}.
Fixing the conjugate momentum above the constant value $J_1=1$ discloses the presence of a critical deformation parameter,  
where the energies of the point and the giant graviton solutions are equal. 
This in turn creates a whole class of solutions (for every value of $J_1$ between 1 and 2 there 
is a critical value of $\Delta$ between 0 and $1/ \sqrt{2}$) 
and the precise relation between $\Delta_{\rm crit}$ and $J_1$ is given in equation \eqref{DvsJ_equation}. 
Below this critical value $\Delta < \Delta_{\rm crit}$ the point graviton has less energy than the giant graviton while above, that is when 
$\Delta > \Delta_{\rm crit}$ it is the giant graviton that is energetically favored.
Notice that until now in the literature (see e.g. \cite{Avramis:2007wb}) the mere presence of a deformation parameter in the 
geometry lifts a degeneracy between the point and the giant graviton solution. The fact that with a Schr\"{o}dinger 
deformation such a degeneracy returns for a finite deformation is an interesting new feature and is related to the 
drastic changes of the dipole-deformation on the geometry, as well as on  the field theory side. 

Besides identifying the two graviton solutions (and motivated by the quantum mechanical expectation of state mixing) 
we focus on cases where the giant graviton is energetically favored with respect to the point graviton and 
construct the instanton solutions tunnelling from the giant to the point graviton. Plotting the total action as a function of the deformation parameter (see e.g. 
figure \ref{fig:ActionVSJ1}) we identify a monotonically decreasing behavior as the deformation parameter increases. This, in turn, means that the transition 
probability increases, something that is totally expected if one carefully examines the shape of the potential in figure \ref{fig:Hamiltonian}. 
Indeed, as the 
deformation parameter increases the potential around the point graviton solution gets flattened and as result the probability 
of tunnelling from the point to the giant graviton increases.  
We give special attention to the case of the critical deformation parameter. In this case the two configurations have the same 
energy and the form of the potential resembles that of the ${\cal N}=4$ case of \cite{Grisaru:2000zn}, but in the presence of 
a deformation. The calculation of the total action is analytic and reveals
 an increased tunnelling probability as the 
deformation parameter increases (or equivalently, due the the relation that appears in equation \eqref{DvsJ_equation}, 
as the conjugate momentum increases).

Finally we focus our attention of the holographic calculation of the three-point correlation functions
of the dilaton modes and two ``heavy" operators, following the analysis that is detailed in \cite{Georgiou:2018zkt}.
In the case at hand, the ``heavy" operator is the giant graviton solution and in particular the class of solutions where 
the deformation parameter acquires the critical value of equation \eqref{DvsJ_equation}. Moreover and in order for the 
classical solution to have the interpretation of a brane tunnelling between two boundary points 
(where the field theory operators will be located), the giant graviton solution in slightly generalized.  
The result of the computation is twofold: From one side we confirm the form of the correlator that is dictated by conformal 
invariance and from the other we specify the scaling function at strong coupling. This is a quantity that is not possible to fix using 
conformal symmetry.

There is a number of interesting directions which have not be explored in the context of Schr\"{o}dinger holography. One of these is related to  instanton calculations in the Schr\"{o}dinger/null-dipole CFT duality. The analogous calculations for the marginal deformation of ${\cal N}=4$ called $\beta$-deformations and for the three parameter deformation of ${\cal N}=4$ \cite{Frolov:2005dj} have been performed in \cite{Georgiou:2006np} and \cite{Durnford:2006nb}, respectively. A second direction would be to focus on the field theory side of the duality and study the gauge invariant operators dual to the giant graviton configurations presented in this work. These should be obtained from the undeformed ones by introducing the appropriate star product among the ${\cal N}=4$ fields which will generate the 
dipole shift along one of the light-cone directions. Finally. it would be interesting to investigate which are the general features that a background should have so that the giant graviton solution is energetically favored compared to the point graviton. Our analysis points towards the direction that deforming the $AdS$ part of the geometry is essential regarding this aspect.



\section*{Acknowledgments}

The work of this project has received funding from the Hellenic Foundation
for Research and Innovation (HFRI) and the General Secretariat for Research and Technology (GSRT), under grant agreement No 15425.





\begin{thebibliography}{99}

\bibitem{McGreevy:2000cw} 
  J.~McGreevy, L.~Susskind and N.~Toumbas,
  ``Invasion of the giant gravitons from Anti-de Sitter space,''
  JHEP {\bf 0006}, 008 (2000)
  [hep-th/0003075].
  
\bibitem{Das:2000st}
S.~R.~Das, A.~Jevicki and S.~D.~Mathur,
``Vibration modes of giant gravitons,''
Phys. Rev. D \textbf{63} (2001), 024013
[arXiv:hep-th/0009019 [hep-th]].
  
  
\bibitem{Grisaru:2000zn} 
  M.~T.~Grisaru, R.~C.~Myers and O.~Tafjord,
  ``SUSY and goliath,''
  JHEP {\bf 0008}, 040 (2000)
  [hep-th/0008015].
  
  
\bibitem{Hashimoto:2000zp} 
  A.~Hashimoto, S.~Hirano and N.~Itzhaki,
  ``Large branes in AdS and their field theory dual,''
  JHEP {\bf 0008}, 051 (2000)
  [hep-th/0008016].
  
 

\bibitem{Pirrone:2006iq} 
  M.~Pirrone,
  ``Giants On Deformed Backgrounds,''
  JHEP {\bf 0612}, 064 (2006)
  [hep-th/0609173].
  
  
\bibitem{Imeroni:2006rb} 
  E.~Imeroni and A.~Naqvi,
  ``Giants and loops in beta-deformed theories,''
  JHEP {\bf 0703}, 034 (2007)
  [hep-th/0612032].
  
  
\bibitem{deMelloKoch:2005jg} 
  R.~de Mello Koch, N.~Ives, J.~Smolic and M.~Smolic,
  ``Unstable giants,''
  Phys.\ Rev.\ D {\bf 73}, 064007 (2006)
  [hep-th/0509007].

\bibitem{Avramis:2007wb} 
  S.~D.~Avramis, K.~Sfetsos and D.~Zoakos,
  ``Complex marginal deformations of D3-brane geometries, their Penrose limits and giant gravitons,''
  Nucl.\ Phys.\ B {\bf 787}, 55 (2007)
  [arXiv:0704.2067 [hep-th]].
  
\bibitem{Huang:2007th}
W.~H.~Huang,
``Thermal Giant Graviton with Non-commutative Dipole Field,''
JHEP \textbf{11} (2007), 015
[arXiv:0709.0320 [hep-th]].

\bibitem{Georgiou:2020qnh}
G.~Georgiou and D.~Zoakos,
``Giant gravitons on the Schrodinger pp-wave geometry,''
JHEP \textbf{03} (2020), 185
[arXiv:2002.05460 [hep-th]].


\bibitem{Maldacena:1997re}
  J.~M.~Maldacena,
  ``The Large N limit of superconformal field theories and supergravity,''
  Int.\ J.\ Theor.\ Phys.\  {\bf 38} (1999) 1113
   [Adv.\ Theor.\ Math.\ Phys.\  {\bf 2} (1998) 231]
  [hep-th/9711200].
  
\bibitem{Maldacena:2008wh}
  J.~Maldacena, D.~Martelli and Y.~Tachikawa,
  ``Comments on string theory backgrounds with non-relativistic conformal symmetry,''
  JHEP {\bf 0810}, 072 (2008)
  [arXiv:0807.1100 [hep-th]].
  
\bibitem{Fuertes:2009ex}
  C.~A.~Fuertes and S.~Moroz,
  ``Correlation functions in the non-relativistic AdS/CFT correspondence,''
  Phys.\ Rev.\ D {\bf 79} (2009) 106004
  [arXiv:0903.1844 [hep-th]].

\bibitem{Volovich:2009yh}
  A.~Volovich and C.~Wen,
  ``Correlation Functions in Non-Relativistic Holography,''
  JHEP {\bf 0905} (2009) 087
  [arXiv:0903.2455 [hep-th]].
  
  \bibitem{Georgiou:2017pvi}
  G.~Georgiou and D.~Zoakos,
  ``Giant magnons and spiky strings in the Schrodinger/dipole-deformed CFT correspondence,''
  JHEP {\bf 1802} (2018) 173
  [arXiv:1712.03091 [hep-th]].
  
\bibitem{Chu:2006ae}
C.~S.~Chu, G.~Georgiou and V.~V.~Khoze,
``Magnons, classical strings and beta-deformations,''
JHEP \textbf{11}, 093 (2006)
[arXiv:hep-th/0606220 [hep-th]].
  
\bibitem{Zoakos:2020gyb}
D.~Zoakos,
``Finite size effects in classical string solutions of the Schrodinger geometry,''
JHEP \textbf{08} (2020), 091
[arXiv:2006.02285 [hep-th]].

  
\bibitem{Ahn:2017bio}
C.~Ahn and P.~Bozhilov,
``Giant magnon-like solution in $Sch_5 \times S^5$,''
Phys. Rev. D \textbf{98}, no.10, 106005 (2018)
[arXiv:1711.09252 [hep-th]].


\bibitem{Golubtsova:2020fpm}
A.~Golubtsova, H.~Dimov, I.~Iliev, M.~Radomirov, R.~C.~Rashkov and T.~Vetsov,
``More on Schrodinger holography,''
JHEP \textbf{08}, 090 (2020)
[arXiv:2004.13802 [hep-th]].


\bibitem{Georgiou:2019lqh}
  G.~Georgiou, K.~Sfetsos and D.~Zoakos,
  ``String theory on the Schrodinger pp-wave background,''
  JHEP {\bf 1908} (2019) 093
  [arXiv:1906.08269 [hep-th]].
  
\bibitem{Guica:2017jmq} 
  M.~Guica, F.~Levkovich-Maslyuk and K.~Zarembo,
  ``Integrability in dipole-deformed $\mathcal{N}=4$ super Yang--Mills,''
  J.\ Phys.\ A {\bf 50}, no. 39, 39 (2017)
  [arXiv:1706.07957 [hep-th]].
  
\bibitem{Ouyang:2017yko}
H.~Ouyang,
``Semiclassical spectrum for BMN string in $Sch_5\times S^5$,''
JHEP \textbf{12} (2017), 126
[arXiv:1709.06844 [hep-th]].


\bibitem{Georgiou:2018zkt}
  G.~Georgiou and D.~Zoakos,
  ``Holographic three-point correlators in the Schrodinger/dipole CFT correspondence,''
  JHEP {\bf 1809} (2018) 026
  [arXiv:1806.08181 [hep-th]].

\bibitem{Dimov:2019koi}
H.~Dimov, M.~Radomirov, R.~C.~Rashkov and T.~Vetsov,
``On pulsating strings in Schrodinger backgrounds,''
JHEP \textbf{10}, 094 (2019)
[arXiv:1903.07444 [hep-th]].

  
  
\bibitem{Golubtsova:2020mjn}
A.~Golubtsova, H.~Dimov, I.~Iliev, M.~Radomirov, R.~C.~Rashkov and T.~Vetsov,
``Pulsating strings in $Schr_5 \times T^{1,1}$ background,''
[arXiv:2007.01665 [hep-th]].


\bibitem{Dimov:2020fzi}
H.~Dimov, M.~Radomirov, R.~C.~Rashkov and T.~Vetsov,
``Holographic Fisher Information Metric in Schr\"odinger Spacetime,''
[arXiv:2009.01123 [hep-th]].


\bibitem{Alishahiha:2003ru}
  M.~Alishahiha and O.~J.~Ganor,
  ``Twisted backgrounds, PP waves and nonlocal field theories,''
  JHEP {\bf 0303}, 006 (2003)
  [hep-th/0301080].


  
\bibitem{Zarembo:2010rr}
  K.~Zarembo,
  ``Holographic three-point functions of semiclassical states,''
  JHEP {\bf 1009} (2010) 030
  [arXiv:1008.1059 [hep-th]].


\bibitem{Costa:2010rz}
  M.~S.~Costa, R.~Monteiro, J.~E.~Santos and D.~Zoakos,
  ``On three-point correlation functions in the gauge/gravity duality,''
  JHEP {\bf 1011} (2010) 141
  [arXiv:1008.1070 [hep-th]].

\bibitem{Georgiou:2010an}
  G.~Georgiou,
  ``Two and three-point correlators of operators dual to folded string solutions at strong coupling,''
  JHEP {\bf 1102}, 046 (2011)
  [arXiv:1011.5181 [hep-th]].

\bibitem{Georgiou:2011qk}
  G.~Georgiou,
  ``SL(2) sector: weak/strong coupling agreement of three-point correlators,''
  JHEP {\bf 1109}, 132 (2011)
  [arXiv:1107.1850 [hep-th]].
  
  
\bibitem{Georgiou:2013ff}
G.~Georgiou, B.~H.~Lee and C.~Park,
``Correlators of massive string states with conserved currents,''
JHEP \textbf{03}, 167 (2013)
[arXiv:1301.5092 [hep-th]].



\bibitem{Bajnok:2016xxu}
  Z.~Bajnok and R.~A.~Janik,
  ``Classical limit of diagonal form factors and HHL correlators,''
  JHEP {\bf 1701}, 063 (2017)
  [arXiv:1607.02830 [hep-th]].
  
  
\bibitem{Bissi:2011dc}
A.~Bissi, C.~Kristjansen, D.~Young and K.~Zoubos,
``Holographic three-point functions of giant gravitons,''
JHEP \textbf{06}, 085 (2011)
[arXiv:1103.4079 [hep-th]].

    
\bibitem{Blau:2009gd}
M.~Blau, J.~Hartong and B.~Rollier,
``Geometry of Schrodinger Space-Times, Global Coordinates, and Harmonic Trapping,''
JHEP \textbf{07}, 027 (2009)
[arXiv:0904.3304 [hep-th]].

\bibitem{Frolov:2005dj}
S.~Frolov,
``Lax pair for strings in Lunin-Maldacena background,''
JHEP \textbf{05}, 069 (2005)
[arXiv:hep-th/0503201 [hep-th]].

\bibitem{Georgiou:2006np}
G.~Georgiou and V.~V.~Khoze,
``Instanton calculations in the beta-deformed AdS/CFT correspondence,''
JHEP \textbf{04}, 049 (2006)
[arXiv:hep-th/0602141 [hep-th]].

\bibitem{Durnford:2006nb}
C.~Durnford, G.~Georgiou and V.~V.~Khoze,
``Instanton test of non-supersymmetric deformations of the $AdS_5 \times S^5$,''
JHEP \textbf{09}, 005 (2006)
[arXiv:hep-th/0606111 [hep-th]].
   
\end{thebibliography}
\end{document}